\documentclass[useAMS,usenatbib,times,a4paper,amssymb]{mn2e}
\usepackage{epsfig,times,amssymb,amsmath,color,verbatim}

\title[CFHTLenS]{CFHTLenS: The Canada-France-Hawaii Telescope Lensing Survey}
\author[C. Heymans et al.]{Catherine Heymans$^{1}$\thanks{heymans@roe.ac.uk}, Ludovic Van Waerbeke$^2$, Lance Miller$^3$, Thomas Erben$^4$, 
\newauthor Hendrik Hildebrandt$^{2,4}$, Henk Hoekstra$^{5,6}$, Thomas D. Kitching$^1$, Yannick Mellier$^7$, 
\newauthor Patrick Simon$^4$, Christopher Bonnett$^{8}$, Jean Coupon$^{9}$, Liping Fu$^{10}$, 
\newauthor Joachim Harnois-D\'{e}raps$^{11,12}$, 
Michael J. Hudson$^{13,14}$, 
Martin Kilbinger$^{15,16,17,7}$,
\newauthor Koenraad Kuijken$^{5}$, 
Barnaby Rowe$^{18,19,20}$, 
Tim Schrabback$^{4,21,5}$,  
Elisabetta Semboloni$^{5}$, 
\newauthor Edo van Uitert$^{4,5}$, Sanaz Vafaei$^{2}$,
Malin Velander$^{3,5}.$\\
\\
$^1$Scottish Universities Physics Alliance, Institute for Astronomy, University of Edinburgh, Royal Observatory, Blackford Hill, Edinburgh, EH9 3HJ, UK. \\ 
$^2$Department of Physics and Astronomy, University of British Columbia, 6224 Agricultural Road, Vancouver, V6T 1Z1, BC, Canada.\\  
$^3$Department of Physics, Oxford University, Keble Road, Oxford OX1 3RH, UK.\\ 
$^4$Argelander Institute for Astronomy, University of Bonn, Auf dem H{\"u}gel 71, 53121 Bonn, Germany.\\
$^5$Leiden Observatory, Leiden University, Niels Bohrweg 2, 2333 CA Leiden, The Netherlands.\\
$^6$Department of Physics and Astronomy, University of Victoria, Victoria, BC V8P 5C2, Canada.\\
$^7$Institut d'Astrophysique de Paris, Universit\'{e} Pierre et Marie Curie - Paris 6, 98 bis Boulevard Arago, F-75014 Paris, France.\\
$^8$Institut de Ciencies de l'Espai, CSIC/IEEC, F. de Ciencies, Torre C5 par-2, Barcelona 08193, Spain.\\
$^9$Institute of Astronomy and Astrophysics, Academia Sinica, P.O. Box 23-141, Taipei 10617, Taiwan.\\
$^{10}$Key Lab for Astrophysics, Shanghai Normal University, 100 Guilin Road, 200234, Shanghai, China. \\
$^{11}$Canadian Institute for Theoretical Astrophysics, University of Toronto, M5S 3H8, Ontario, Canada.\\
$^{12}$Department of Physics, University of Toronto, M5S 1A7, Ontario, Canada.\\
$^{13}$Department of Physics and Astronomy, University of Waterloo, Waterloo, ON, N2L 3G1, Canada.\\
$^{14}$Perimeter Institute for Theoretical Physics, 31 Caroline Street N, Waterloo, ON, N2L 1Y5, Canada.\\
$^{15}$CEA Saclay, Service d'Astrophysique (SAp), Orme des Merisiers, B\^at 709, F-91191 Gif-sur-Yvette, France.\\
$^{16}$Excellence Cluster Universe, Boltzmannstr. 2, D-85748 Garching, Germany.\\
$^{17}$Universit\"ats-Sternwarte, Ludwig-Maximillians-Universit\"at M\"unchen, Scheinerstr.~1, 81679 M\"unchen, Germany.\\
$^{18}$Department of Physics and Astronomy, University College London, Gower Street, London WC1E 6BT, UK.\\
$^{19}$Jet Propulsion Laboratory, California Institute of Technology, 4800 Oak Grove Drive, Pasadena CA 91109, USA.\\
$^{20}$California Institute of Technology, 1200 E California Boulevard, Pasadena CA 91125, USA.\\
$^{21}$Kavli Institute for Particle Astrophysics and Cosmology, Stanford University, 382 Via Pueblo Mall, Stanford, CA 94305-4060, USA.\\
}
\newcommand{\be}{\begin{equation}}  \newcommand{\ee}{\end{equation}}
  \newcommand{\ba}{\begin{eqnarray}}
\newcommand{\bm}[1]{\mbox{\boldmath{$#1$}}}   

\renewcommand{\d}[0]{{\rm d}}

\newcommand{\Ave}[1]{\Big\langle #1 \Big\rangle}

\def\gs{\mathrel{\raise1.16pt\hbox{$>$}\kern-7.0pt %
\lower3.06pt\hbox{{$\scriptstyle \sim$}}}}         %
\def\ls{\mathrel{\raise1.16pt\hbox{$<$}\kern-7.0pt %
\lower3.06pt\hbox{{$\scriptstyle \sim$}}}}         %

\begin{document}

\maketitle

\begin{abstract}
We present the Canada-France-Hawaii Telescope Lensing Survey (CFHTLenS) that accurately determines a weak gravitational lensing signal from the full 154 square degrees of deep multi-colour data obtained by the CFHT Legacy Survey.  Weak gravitational lensing by large-scale structure is widely recognised as one of the most powerful but technically challenging probes of cosmology.  We outline the CFHTLenS analysis pipeline, describing how and why every step of the chain from the raw pixel data to the lensing shear and photometric redshift measurement has been revised and improved compared to previous analyses of a subset of the same data.  We present a novel method to identify data which contributes a non-negligible contamination to our sample and quantify the required level of calibration for the survey.  Through a series of cosmology-insensitive tests we demonstrate the robustness of the resulting cosmic shear signal, presenting a science-ready shear and photometric redshift catalogue for future exploitation.   

\vspace{0.7cm}

\end{abstract}


\section{Introduction}
Our understanding of the Universe has grown rapidly over the past decade.  Heralded as the era of high precision cosmology, multiple independent and diverse observations all point to a Universe dominated by dark energy and dark matter.  The concordant cosmology derived from these observations accurately determines the composition of the Universe \citep[see review article][and references therein]{Weinberg} and the highest priority is now to understand the different phenomena that comprise what is often referred to as the Dark Universe.  

Weak gravitational lensing is a unique tool for cosmology which directly probes the mass distribution of matter in the Universe independent of its state or nature.  As light propagates through the Universe its path becomes deflected by the gravitational potential of the large-scale structures of matter, with the consequence that distant galaxy images are observed to be weakly and coherently distorted.  This observation can be directly related to the underlying matter power spectrum of the Universe \citep[see for example][]{TS10}
and can pinpoint where matter is and how much of it there is \citep{MasseyNat,Hey08}.  Compared to other probes of cosmology, weak lensing is particularly interesting as it provides a direct measurement of the growth of large-scale structures in the Universe allowing us to test the fundamental and alternative gravity theories suggested to explain the dark energy in the Universe \citep{Reyes,Simpson}.  

Weak lensing by large-scale structure is widely recognised as one of the most powerful but technically challenging probes of cosmology.  Since its first detection \citep{BRE,vWb00,Witt} advances in technology and deep-wide survey observations have been paralleled by significant community efforts in algorithm development  in order to overcome the challenges of this non-trivial observation \citep{STEP1,STEP2,GREAT08,PHAT,GREAT10,MDM}.  The measurement requires the detection of percent level shear distortions imprinted on the images of distant galaxies by cosmological lensing in the presence of temporal and spatially varying $\sim 10\%$ distortions caused by the atmosphere, telescope and detector.  The growth in precision of lensing surveys over the past decade has required an increasing accuracy in the understanding of the origins of the distortions and the impact of data reduction methods on our shear measurement algorithms.  These local effects are encompassed in the point spread function (PSF) as measured from images of stellar objects in the survey.

In this paper we present the Canada-France-Hawaii Telescope Lensing Survey (CFHTLenS) that accurately measures weak gravitational lensing from the deep multi-colour data obtained as part of the CFHT Legacy Survey (CFHTLS).  This survey spans 154 square degrees in five optical bands ${\it u^*g'r'i'z'}$ with a $5\sigma$ point source limiting magnitude in the $i'$-band of $i'_{AB}\sim 25.5$.  CFHTLenS incorporates data from the main Wide survey, Deep survey, the astrometric pre-imaging and photometric calibration post-imaging components of the CFHTLS which completed observations in early 2009.    The first weak lensing analysis of the CFHTLS-Wide \citep{HH06} analysed 31 square degrees of single band $i'$ data, showing the high quality imaging obtained by the then newly commissioned one square degree field of view MegaCam imager on the 3.6m CFHT.  This conservative analysis selected regions of the data far from the boundaries of the individual CCD chips in the MegaCam imager, omitting one-third of the data.  This strategy was used in order to circumvent issues associated with the stacked combination of varied PSFs from different chips in the 7 dithered exposures in these regions.  This first analysis was followed by \citet{Fu08} where the 57 square degrees of the survey analysed had, for the first time, sufficient statistical accuracy to reveal hints of systematic errors in the measured shear signal on large scales.  Significant variations of the shear signal between individual MegaCam pointings were then uncovered \citep{MK09}.  This hinted at a problem with understanding the PSF even though a similarly conservative masking scheme had been applied to reject the problematic image regions excised in the \citet{HH06} analysis.  In addition a problem with the redshift-scaling of the two-point shear correlation function soon became apparent when the data used in \citet{Fu08} were combined with photometric redshift measurements from \citet{Coupon09}, \citep[see][for more details]{MK09}.   The CFHTLenS collaboration formed to develop new techniques and find a solution to reduce these systematic errors.  

In this paper we start in Section~\ref{sec:pipe} by outlining the CFHTLenS analysis pipeline and the importance of working with individual exposures rather than stacked images.  Parts of the pipeline are presented in much finer detail in \citet[][pixel analysis]{Erben12}, \citet[][photometry and redshift analysis]{HH12}, and \citet[][shear analysis]{Miller2012}.  We then detail the methodology behind our cosmology-insensitive and quantitive systematic error analysis in Section~\ref{sec:method} and present the results of that analysis, calibrating and selecting a clean data sample, in Section~\ref{sec:res}.  We investigate the impact of the calibration and data selection on the two-point shear correlation function in Section~\ref{sec:2pt}, comparing our systematic error analysis to alternative tests advocated by previous weak lensing analyses.  We additionally show the robustness of the measurements by performing a final demonstration that the shear signal is not subject to redshift-dependent biases in Section~\ref{sec:zscale}, using a cosmology-insensitive galaxy-galaxy lensing test.  Finally  we conclude in Section~\ref{sec:conc}.   Throughout this paper the presented errors show the $1\sigma$ confidence regions as computed through bootstrap analyses of the data, unless stated otherwise.

\begin{table*}
\caption{Comparison of the key stages in the lensing analysis pipelines used by CFHTLenS and in an earlier CFHTLS analysis by \citet{Fu08}.  Note that the earlier CFHTLS pipeline is a good example of the standard methods that have been used by the majority of weak lensing analyses to date (c. 2012).} 
\centering      
\begin{tabular}{l l l}  
\hline\hline                        
Pipeline Stage & \citet{Fu08} CFHTLS Pipeline & CFHTLenS Pipeline \\ [0.5ex] 
\hline                    
Astrometry reference catalogue & USNO-B1 & 2MASS \& SDSS DR7 \\
Cosmic Ray Rejection method & Median pixel count in stack & {\sc SExtractor} neural-network detection  \\
& & refined to prevent the misclassification of stellar cores  \\
Photometry measured on & Re-pixelised median stacks & Re-pixelised mean stacks with a gaussianised PSF \\
Star-galaxy size-mag selection & Averaged over each field of view (Pipeline I) &  Chip-by-chip selection  \\ 
& Chip-by-chip selection (Pipeline II) &  plus additional colour selection \\ 
Redshift distribution & Extrapolation from CFHTLS-Deep fields & From {\sc BPZ} photometric redshift measurements of each galaxy  \\ 
Shape Measurement method& {\sc KSB+} & {\em lens}fit\\ 
Shapes measured on & Re-pixelised median stacks & Individual exposures with no re-pixelisation \\
Systematic error analysis & Averaged results for full survey & Exposure level tests on individual fields  \\ [1ex]       
\hline     
\end{tabular} 
\label{tab:comparison}  
\end{table*} 

\section{The CFHTLenS Analysis Pipeline}
\label{sec:pipe}
The CFHTLenS collaboration set out to robustly test every stage of weak lensing data analysis from the raw pixel data through data reduction to object detection, selection, photometry, shear and redshift estimates and finally systematic error analysis.  For each step in the analysis chain, multiple methods were tested and rejected or improved such that the final CFHTLenS pipeline used in this analysis re-wrote every single stage of the data analysis used in earlier CFHTLS analyses, and indeed the majority of all previous weak lensing analyses.   We argue that there was not a single distinct reason why earlier CFHTLS lensing analyses suffered from systematic errors by discussing a series of potential successive sources of systematic error that could have accumulated throughout earlier analyses.  We detail each step in the analysis chain in this section for users and developers of weak lensing analysis pipelines.   For those readers interested in the main changes a summary is provided in Table~\ref{tab:comparison} which compares the key stages in the lensing analysis pipelines used by CFHTLenS and in an earlier CFHTLS analysis by \citet{Fu08}.  Note that the \citet{Fu08} pipeline is a good example of the standard methods that have been used by the majority of weak lensing analyses to date (c. 2012).  One notable exception is \citet{BJ02} who advocate the weak lensing analysis of individual exposure data rather than stacks, by averaging the shear measurements from multiple exposures on a catalogue level.  The extension of this exposure-level analysis argument to the more optimal, simultaneous, joint model-fitting analysis of multiple exposures, has been one of the most important revisions in our analysis.

\subsection{CFHTLS Data}

The CFHTLS-Wide data spans four distinct contiguous fields:  W1 ($\sim 63.8$ square degrees), W2 ($\sim 22.6$ square degrees), W3 ($\sim 44.2$ square degrees) and W4 ($\sim 23.3$ square degrees). The survey strategy was optimised for the study of weak gravitational lensing by reserving the observing periods with seeing better than $\sim 0.8$ arcsec for the primary lensing $i'$-band imaging.  The other $u^*g'r'z'$ bands were imaged in the poorer seeing conditions.  A detailed report of the full CFHTLS Deep and Wide surveys can be found in the TERAPIX CFHTLS T0006 release document\footnote{The CFHTLS T0006 Release Document:  {\it http://terapix.iap.fr/cplt/T0006-doc.pdf}}.  All CFHT MegaCam images are initially processed using the {\sc Elixir} software at the Canadian Astronomical Data Centre \citep{Elixir}.  We use the {\sc Elixir} instrument calibrations and de-trended archived data as a starting point for the CFHTLenS project.  

\subsection{Data Reduction with {\sc THELI}}
\label{sec:cr}
The CFHTLenS data analysis pipeline starts with the public {\sc THELI} data reduction pipeline designed to produce lensing quality data \citep{THELI} and first applied to the CFHTLS in \citet{Erben09}.    We produce co-added weighted mean stacks for object detection and photometry and use single exposure $i'$-band images for the lensing analysis.  Relevant pixel quality information in the form of weights is produced for each image, both for the stacks and the individual exposures.  

Improvement of the cosmic ray rejection algorithm was one of the key developments in {\sc THELI} for CFHTLenS.  We tested the robustness of the neural-network procedure implemented in the {\sc SExtractor} software package \citep{SExtractor} to identify cosmic ray hits using the default MegaCam cosmic ray filter created by the {\sc EyE} (Enhance your Extraction) software package \citep{EyE}.  We found this set-up to reliably identify cosmic ray hits but at the expense of the misclassification of the bright cores of stars in images with a seeing better than $\sim 0.7$ arcsec.  In images with a seeing of $< 0.6$ arcsec, nearly
all stars brighter than $i'_{\rm AB}\approx 19$ had cores misclassified as a cosmic ray defect.  This misclassification is particularly problematic for lensing analyses that analyse individual exposures as a non-negligible fraction of stars used for modeling the PSF are then rejected by the cosmic ray mask.  In addition this rejection may not be random, for example the neural-network procedure may preferentially reject pixels in the cores of the stars with the highest Strehl ratio thus artificially reducing the average Strehl ratio of the PSF model in that region.  For lensing analyses that use co-added stacks, this cosmic ray misclassification is also an issue when using mean co-added images of the dithered exposures.  The misclassified exposure level pixels are, by definition, the brightest at that location within the stack.  When the cosmic ray mask removes these misclassified pixels from the co-addition, the centres of stars therefore artificially lack flux which produces an error in the PSF model.     

In earlier analyses of CFHTLS, concerns about cosmic ray rejection were circumvented by using a median co-added image of the dithered exposures, even though this method does not maximise the signal-to-noise of the final co-added image and can only be applied to images with enough exposures.   In this case whilst the misclassification of cosmic rays is likely no longer an issue, a more subtle PSF effect is at play.  As the median is a non-linear statistic of the individual pixel values, it destroys the convolutional relationship between the stars and galaxies in the stacked images such that the PSF from the stars differs from the PSF experienced by the galaxies.  This can be illustrated by considering the case of four dithered exposures, three with similar PSF sizes and ellipticities and one exposure with a PSF with fainter extended wings.  In the median co-added image, the PSF modeled from the stellar objects will match the PSF in the first three exposures, as the fourth exposure is rejected at the location of the stars by the median operation on the image.  The PSF as seen by the fainter extended galaxies will however include the effects of the extended wing PSF from the fourth exposure.    This is because pixel noise and the broader smoothing from the extended fourth PSF exposure can combine in such a way that the fourth exposure determines the final pixel count at some locations across the median co-added galaxy image.  In the CFHTLS there are typically 7 exposures per image with significant variation of the PSF between exposures such that the previous use of median co-added images could well have contributed to the systematic error found in earlier CFHTLS analyses.

For the current work we refined the standard procedures for cosmic ray flagging by identifying cosmic rays in a two-stage process.  First we identify cosmic rays using the neural-network procedure in the {\sc SExtractor} software package, as described above.  We then extract a catalogue of bright sources from the data using {\sc SExtractor} with a high detection threshold set, requiring more than 10 pixels to be connected with counts above $10 \sigma$.  Candidate unsaturated stars on the image are then selected using the automated {\sc THELI} routine which locates the stellar locus in the size-magnitude frame.  We then perform a standard PSF analysis to clean the bright candidate star catalogue using the `Kaiser, Squires and Broadhurst'  method \citep[][hereafter {\sc KSB+}]{KSB}. This consists of measuring second order brightness moments and performing a two-dimensional second order polynomial iterative fit to the PSF anisotropy with outliers removed to obtain a clean bright star catalogue. As this particular stellar sample is bright, this method is very effective at selecting stars.  All bright stellar sources that remain in the sample in the final fit are then freed from any cosmic ray masking in the first step.  This procedure was found to produce a clean and complete cosmic ray mask that left untouched the bright end of the stellar branch required for the more thorough CFHTLenS star-galaxy classification and subsequent PSF analyses (see Section~\ref{subsec:photom}).   The method will, however, allow through the very rare cases of real cosmic ray defects at the location of stars.  In these cases, the stellar objects will likely be flagged as unusual in the colour-colour stellar selection stage that follows, as described in section~\ref{subsec:photom}.  Further properties of this method are detailed in \citet{Erben12}.

In addition to advances in cosmic ray identification, {\sc THELI} was updated to perform photometric and astrometric calibrations over each survey patch (W1,W2, W3 and W4) aided by the sparse astrometric pre-imaging data and photometric calibration post-imaging data (programs 08AL99 and 08BL99), in contrast to earlier calibration analyses on a square degree MegaCam pointing basis.  The resulting field-to-field RMS uncertainty in relative photometry is $\sigma \sim 0.01-0.03$ magnitude in all passbands. The uniform internal astrometry has a field-to-field RMS error $\sigma \sim 0.02''$.  Significant effort was invested in testing the impact of using different reference catalogues (2MASS with SDSS-DR7 was found to be the most accurate) in addition to the robustness of the {\sc Scamp} astrometric software \citep{scamp} and {\sc Swarp} re-pixelisation software \citep{swarp}, and the impact of the astrometric distortion correction, interpolation and re-pixelisation on the PSF and galaxy shapes using simulated data.  The conclusion of this work was that whilst these methods were excellent for astrometry they were not sufficiently accurate for shape measurement, particularly if the imaging was undersampled.    For this reason we do not interpolate or re-pixelate the data used for our lensing analysis.  Instead we apply the derived astrometric distortion model to the galaxy models when model fitting to the data \citep{Miller2012}.  Finally, the automated {\sc THELI} masking routine was applied to the data to identify saturated stars, satellite trails, ghosts and other artifacts.  These masks were individually expected, verified and improved manually, and the importance of this manual inspection is investigated in Section~\ref{sec:masks}.  A more detailed description of the {\sc THELI} analysis in CFHTLenS is presented in \citet{Erben12}.

\subsection{Object Selection, PSF characterisation and Photometric Redshifts with Gaussianised Photometry and {\sc BPZ}}
\label{subsec:photom}

The next stage in the CFHTLenS data analysis pipeline is object detection and classification.   We use the {\sc SExtractor} software \citep{SExtractor} to detect sources in the $i'$-band stacks.  
This initial catalogue forms the basis for the shape and photometric measurements that follow.  Using  the $i'$-band as the primary detection image preserves the depth of the co-added data for our lensing population and it is this complete initial catalogue that is carried through the full pipeline with flags and weights added to indicate the quality of the derived photometric and shape parameters.

We manually select stars in the size-magnitude plane.  In this plane the stellar locus at bright magnitudes and fixed constant size, governed by the PSF, can be readily identified from the galaxy population.  Stars are selected down to 0.2 magnitudes brighter than the magnitude limit where the stellar locus and galaxy population merge.  We found the manual selection was necessary to select a sufficient number of reliable stars at magnitudes fainter than the bright stellar sample that was automatically selected by the {\sc THELI} routine and used to improve the cosmic ray rejection algorithm (see Section~\ref{sec:cr}).  The manual selection is performed on a chip-by-chip basis as the stellar size varies significantly across the field of view.  If the star-galaxy separation is made over the whole field this can result in truncation in the stellar density at the extremes of the stellar locus in the size-magnitude plane.  This would then result in a lack of stars in the corresponding regions in the field of view and hence a poor PSF model in those regions.    This type of full field selection was part of the main Pipeline I CFHTLS analysis of \citet{Fu08}, with a manual chip-based selection reserved only for those fields which exhibited an unusually wide stellar locus scatter.   The upcoming launch of the ESA-Gaia mission\footnote{Gaia: gaia.esa.int} to survey over a billion stars in our galaxy will mean that this manual stellar selection stage will not be required in the future.

For lensing studies we require a pure and representative star catalogue across the field of view that does not select against a particular PSF size or shape.  Provided the catalogue is representative it does not need to be complete.   To ensure the purity of the sample selected in the size-magnitude plane we perform an additional colour analysis.  We first construct a 4D colour space; $g'-r'$, $r'-i'$, $i'-z'$ and $g'-z'$, and determine the distribution of stellar candidates in this space.  A stellar candidate is confirmed to be a star when more than five percent of the total number of candidates lie within a distance of $\sim 0.5$ magnitudes from its location in the 4D space.  We reject any object that lies below this threshold density, typically rejecting about ten percent of the original, manually selected, stellar candidate list.   

An accurate spatially varying pixelised PSF model in each exposure of the lensing $i'$-band data was created using the pure star catalogue for use in the lensing analysis. Each PSF pixel value was modeled using a two-dimensional third-order polynomial function of position in the camera field of view. To allow for the discontinuities in the PSF across the boundaries between CCDs, some coefficients of the polynomial are allowed to vary between CCDs \citep[see][for details]{Miller2012}.  Note that the width of the CFHT $i'$-band filter is sufficiently narrow that we can assume the PSF is independent of the star colour \citep{Cypriano10,Voigt}, which is shown to be a good assumption for the $i'$-band in \citet{SNLS}.  We also ignore the differing affects of atmospheric dispersion on the PSF of stars and galaxies. The wavelength dependence of the refractive index of the atmosphere will induce a spectrum dependent elongation of the object.  For the low airmass $i'$-band observations of the CFHTLS, however, the difference in this elongation for stars and galaxies is shown to be negligible \citep{K00}.   We use unweighted quadrupole moment measures of the resulting high signal-to-noise pixelised PSF models to calculate a measure of the PSF ellipticity $e^\star$ at the location of each object \citep[see equation 3 in][]{STEP1}.  These PSF ellipticity estimates are only used in the systematics analysis detailed in Section~\ref{sec:method}.

The pure star catalogue is also used to measure accurate multi-band photometry by constructing spatially varying kernels in ${\it u^*g'r'i'z}'$ which are used to gaussianise the PSF in the mean co-added image of each band.  This data manipulation is used only for photometry measurements, and results in a circular Gaussian PSF that is constant between the bands and across each field of view  \citep[see][for more details on PSF gaussianisation for photometry]{HH12}.  Object colours are determined by running {\sc SExtractor} in dual-image mode on the homogeneous gaussianised ${\it u^*g'r'i'z'}$ images, and this photometry is then analysed using the Bayesian Photometric Redshift Code \citep[][{\sc BPZ}]{BPZ} with a modified galaxy template set, stellar template set and a modified prior as detailed in \citet{HH12}.   

So far we have discussed how to obtain a pure stellar catalogue but for our science goals we also need a pure source galaxy catalogue, which again need not necessarily be complete.  To avoid any potential bias introduced by the manual star selection we use {\em lens}fit \citep{Miller2012} to measure shapes for every object detected with $i'_{\rm AB} <24.7$.  The rationale for this {\sc SExtractor} MAG\_AUTO measured magnitude limit is described below.  Objects are fit with galaxy and PSF models and any unresolved or stellar objects are assigned a weight of zero in the final shape catalogue.    The purity of the resulting galaxy catalogue is confirmed by comparing with a photometry only analysis.  We use {\sc BPZ} to compare the multi-band photometry of each object with galaxy and stellar templates and classify stars and galaxies based on the maximum likelihood found for each object template and the size of the object relative to the sizes of stars as defined in our initial stellar selection  \citep{HH12}.  We find these two methods agree very well in creating a pure galaxy sample with less than 1\% of objects having a different object classification in each method.  These differences occur at faint magnitudes where the low weight in the lensing analysis means that if these objects are truly stars, they would have a negligible impact on the lensing signal.   

The input pure galaxy catalogue is also required to be free of any shape-dependent selection bias.  This could arise, for example, if there is a preference to extract galaxies oriented in the same direction as the PSF \citep{K00} or preferentially extract more circular galaxies such as those that are anti-correlated with the lensing shear \citep{Hirata}.  \citet{STEP1} concluded that for the {\sc SExtractor} algorithm used in this analysis, selection bias was consistent with zero change in the mean ellipticity of the population, and at worse introduced a very weak sub-percent level error on the shear measurement.  We therefore do not consider object detection selection bias any further in our analysis.

The accuracy of the resulting photometric redshifts for the bright end of our pure galaxy sample can be established by comparing photometric redshift estimates to spectroscopic redshifts in the field \citep{HH12}. The public spectroscopic redshifts available in our fields come from the Sloan Digital Sky Survey \citep[SDSS,][]{SDSS-DR7}, the VIMOS VLT deep survey \citep[VVDS,][]{VVDS05} and the DEEP2 galaxy redshift survey \citep{DEEP2}.  For these surveys, the spectroscopic completeness is a strong function of magnitude with the faintest limits yielding $\sim100$ spectra at $i'_{AB} \sim 24.7$, according to our CFHTLenS magnitude estimates.   We choose to limit all our shape and photometric redshift analysis to a magnitude $i'_{AB} \sim 24.7$ even though the imaging data permits one to detect and determine photometry and shapes for objects at slightly deeper magnitudes.  The rationale here is that all CFHTLenS science cases require accurate photometric redshift information and we wish to avoid extrapolating the redshift measurements into a regime with essentially no spectroscopic redshifts. That said, the completeness of the spectroscopy degrades significantly for $i'_{AB}>22.5$.  We therefore cannot trust that the accuracy predicted from a standard photometric-spectroscopic redshift comparison at magnitudes fainter than $i'_{AB}>22.5$ is representative of the full galaxy sample at these magnitudes. In order to address this issue we implement a rigorous analysis to test the accuracy of the photometric redshifts to the faint $i'_{AB} \sim 24.7$ magnitude limit in \citet{Benjamin2012}. In this analysis we split the full galaxy sample into six photometric redshift bins and compare the redshift distributions as determined from the sum of the {\sc BPZ} photometric redshift probability distributions, a cross-bin angular galaxy clustering technique \citep{JB10} and the COSMOS-30 redshifts \citep{COSMOS30}. From this analysis we conclude that, when limited to the photometric redshift range $0.2<z<1.3$, our photometric redshift error distribution is sufficiently well characterized by the measured redshift probability distributions, for the main science goals of the survey \citep{Benjamin2012}.  We can therefore incorporate the
redshift probability distributions in scientific analyses of CFHTLenS in order to account for catastrophic outliers and other redshift errors. Finally, with a magnitude limit fainter than $i'_{AB}=24.7$ objects are detected in the lensing $i'$-band at signal-to-noise ratios less than $7\sigma$ which is a regime where even the best shape measurement methods become very strongly biased \citep{GREAT10}.

\subsection{Shape Measurement Method Selection}
In the early stages of the analysis, the CFHTLenS team tested a variety of different shape measurement methods including two different versions of {\sc KSB+} \citep[most recently used in][]{TS10,Fu08}, three different versions of {\sc Shapelets} \citep{KKshapes,shapelets07,Velander} and the model fitting method {\em lens}fit \citep{Lensfit1,Lensfit2,Miller2012}.    Out of all the measurement algorithms tested by CFHTLenS, only {\em lens}fit had the capability to simultaneously analyse data on an individual exposure level rather than on higher signal-to-noise stacked data where the PSF of the stack is a complex combination of PSFs from different regions of the camera.    During the early multi-method comparison analyses, it became apparent how important this {\em lens}fit capability was.
In principle, all other tested algorithms could have also operated on individual exposures and averaged the results on a catalogue level, as advocated by \citet{BJ02}.    As these methods apply signal-to-noise cuts, however, \citep[see for example Table A1 in][]{STEP1},  this would have yielded a low galaxy number density in comparison to the stacked data analysis.  The signal-to-noise is typically decreased by roughly 40\% on the exposure image compared to the stack.   Additionally, all shape measurement methods are expected to be subject to noise bias which increases as the signal-to-noise of the object decreases \citep{MelchiorViola}.  Averaging catalogues measured from individual lower signal-to-noise exposures is therefore expected to be more biased than a single pass higher signal-to-noise simultaneous analysis of the exposures, as carried out by {\em lens}fit. 

We initially focused on finding a method which did not produce the significant variations of the large-scale shear signal between individual MegaCam pointings as discussed in \citet{MK09}.  We rapidly came to the conclusion that only the model fitting {\em lens}fit analysis could produce a robust result.   In the case of {\sc KSB+}, the PSF is assumed to be a small but highly anisotropic distortion convolved with a large circularly symmetric seeing disk \citep{KSB}.  In the CFHTLenS stacks the PSF does not meet these assumptions so it is not surprising that the {\sc KSB+} method fails on the stacked data, where the PSF varies significantly between exposures. In addition some of the CFHT PSF distortion arises from coma which also cannot be modeled by the assumed {\sc KSB+} PSF profile.  For the {\sc Shapelet} methods we found the signal-to-noise of the shear measurements to be very low in comparison to the other shape measurement methods tested.  This is in contrast to a successful application of the {\sc Shapelet} method to space-based observations where the higher order moment analysis is able to take advantage of the additional resolution \citep{Velander}.  The \citet{KKshapes} {\sc Shapelet} method is, however, used in the CFHTLenS analysis for gaussianising the PSFs for optimal photometry \citep[see Section~\ref{subsec:photom} and][]{HH12}.  In the case of multi-band photometry the gain in computational speed that results from the analytical convolutions that are possible within the {\sc Shapelet} formalism means using {\sc Shapelets} for gaussianisation is preferable to using the slower, but potentially more exact, pixel based model of the PSF as used by {\em lens}fit.  Work is ongoing to see whether the {\sc Shapelets} PSF gaussianisation method can improve the performance of {\sc KSB+}.

It is natural to ask why systematic errors have not been apparent when the methods CFHTLenS tested have previously been tested in the following blind image simulation analysis challenges: \citet{STEP1,STEP2,GREAT08,GREAT10}.   This is particularly relevant to ask as {\sc KSB+} consistently performs well in these challenges. The answer to this lies in the fact that the errors discussed above are a consequence of features in the data that have not been present in the image simulation challenges.  These challenges do not contain astrometric distortions and either have a constant PSF, or known PSF models and stellar locations, with typically low PSF ellipticities.  It is interesting to note that the only shape measurement challenge to simulate the relatively strong 10\%-level PSF distortions that are typical in CFHT MegaCam imaging found significant errors for {\sc KSB+} in this regime \citep{STEP2}.  Finally, low-signal-to-noise multiple dithered exposures have only recently been simulated and tested for use in combination in \citet{Miller2012}, in addition to \citet{GREAT10} which presented the first deep multi-epoch image simulations of non-dithered exposures.  These new simulations are the first to test the difficulty of optimally co-adding exposures in analyses of multiple exposure data such as CFHTLenS.  Even with this advance, features of multiple exposure data such as gaps and discontinuities in coverage and the requirement for interpolation of data with an astrometric distortion remain to be tested in future image simulation challenges, in addition to a more realistic set of galaxy models\footnote{See for example the GREAT3 challenge: www.great3challenge.info}.

\subsection{Shape Measurement with {\em lens}fit}
CFHTLenS is the first weak lensing survey to apply the {\em lens}fit model fitting method and as such there have been many key developments of the algorithm for CFHTLenS.     The method performs a Bayesian model fit to the data, varying galaxy ellipticity and size and marginalising over centroid position.  It uses a forward convolution process, convolving the galaxy models with the PSF to calculate the posterior probability of the model given the data.  A galaxy is then assigned an ellipticity, or shear estimate, $\epsilon$, estimated from the mean likelihood of the model posterior probability, marginalized over galaxy size, centroid and bulge fraction.  An inverse variance weight $w$ is also assigned which is given by the variance of the ellipticity likelihood surface and the variance of the ellipticity distribution of the galaxy population \citep[see][for more details]{Miller2012}.
A summary of the CFHTLenS improvements to the algorithm includes using two-component galaxy models, a new size prior derived from high resolution Hubble Space Telescope data, a new ellipticity prior derived from well resolved galaxies in the SDSS, the application of the astrometric distortion correction to the galaxy model rather than re-pixelising the data, and the simultaneous joint analysis of single dithered exposures rather than the analysis of a stack.  These developments are presented in \citet{Miller2012} along with details of the verification of calibration requirements as determined from the analysis of a new suite of image simulations of dithered low-signal-to-noise exposures.

\subsection{Summary}

Once the {\sc THELI} data analysis, object selection, redshift estimation with {\sc BPZ} and shear estimation with {\em lens}fit are complete, we obtain a galaxy catalogue containing a shear measurement with an inverse variance weight $w$ and a photometric redshift estimate with a probability distribution $P(z)$.    
The number density of galaxies with shear and redshift data is 17 galaxies per square arcmin.  The effective weighted galaxy number density that is useful for a lensing analysis is given by
\be
n_{\rm eff} = \frac{1}{\Omega}\frac{(\sum w_i)^2}{\sum w_i^2} \, ,
\ee
where $\Omega$ is the total area of the survey excluding masked regions, and the sum over weights $w_i$ is taken over all galaxies in the survey.  
We find $n_{\rm eff} =14$ galaxies per square arcmin for the full sample.  We choose, however, to use only those galaxies in our analyses with a photometric redshift estimate between $0.2 < z < 1.3$ \citep[see section~\ref{subsec:photom} and][]{Benjamin2012}.  The deep imaging results in a weighted mean redshift for this sample of $\bar{z} = 0.75$, and a weighted median redshift for this sample of $z_m = 0.7$, as determined from the weighted sum of the $P(z)$.   The effective weighted galaxy number density, in this redshift range, is $n_{\rm eff} =11$ galaxies per square arcmin. 
This photometric redshift selection ensures relatively accurate photometric redshifts across the survey with an average scatter\footnote{The scatter, $\sigma_z$, on the photometric redshifts, $z_{\rm phot}$, is calculated from a comparison with the VVDS and DEEP2 spectroscopic redshifts, $z_{\rm spec}$.  The quoted  $\sigma_z$ is given by
the standard deviation around the mean of $\Delta z = (z_{\rm phot} - z_{\rm spec} )/ (1+z_{\rm spec} )$, after outliers with $|\Delta z| < 0.15$ are removed.  See \citet{HH12} for further details.}  $\sigma_z \sim 0.04(1+z)$ and an average catastrophic outlier rate below 4\%.  Across the selected redshift range, the scatter is always less than $\sigma_z < 0.055(1+z)$ and the catastrophic outliers rate is always less than 11\% \citep{HH12}.  As detailed in \citet{Miller2012}, and discussed further in Section~\ref{sec:empcor}, any calibration corrections applied to the shear measurement are less than 6\% on average.   The task is now to verify the quality of these catalogues and select the data that meets the criterion of negligible systematic errors for a cosmic shear analysis.  The resulting error analysis and field selection is also directly relevant for galaxy, group and cluster lensing analyses of dark matter haloes.  The requirements on systematics for these analyses, however, are typically less stringent as a result of the azimuthal averaging of the lensing signal which reduces the impact of any PSF residuals.  We dedicate the rest of the paper to the derivation, application and validation of a set of systematics criteria to the CFHTLenS data.

To conclude this section we refer the reader back to Table~\ref{tab:comparison} which compares the differences between the key stages in the lensing analysis pipeline described above and the pipeline used in an earlier CFHTLS analysis by \citet{Fu08}, illustrating how every core stage of the pipeline has been re-written for CFHTLenS.  This was necessary as every stage of the standard pipeline used in previous analyses could have introduced low-level systematic errors, with the most important errors coming from the analysis of median stacked data in comparison to individual exposures.

\section{Methods: quantitative systematic error analysis}
\label{sec:method}

The observational measurement of weak gravitational lensing is a challenging task, with the cosmological shear signal $\gamma$ that we wish to extract being roughly an order of magnitude below the atmospheric and telescope distortion.  These artificial sources of distortion are encapsulated in the PSF.  Astrometric camera shear distortion is then applied after the PSF convolution \citep[see][for further discussion on these different types of distortions]{Miller2012}.   We detail the {\em lens}fit shear measurement method in \citet{Miller2012}
and verify and calibrate the robustness of the method on an extensive suite of realistic image simulations.  Whilst this successful demonstration on image simulations is very necessary, it is, however, not sufficient to then conclude the method will also yield an unbiased measure in its application to data.  An example failure could result from any data-related feature not included in the image simulations that would result, for example, in an inaccuracy in the PSF model \citep{HH04,vWb05, Rowe10,HeyRowe}.   In this section we therefore develop a procedure to determine the level of any residual distortions in the shape catalogues which result from an incomplete correction for the true PSF.   In order to distinguish any residual distortions from random noise correlations we need to construct an estimator that takes into account the different sources of noise in our analysis for which we require and develop a set of realistic simulated mock catalogues.  

Throughout the paper when we refer to galaxy ellipticity $\epsilon$, galaxy shear $\gamma$, PSF ellipticity $e^\star$ or noise $\eta$ on any of these shape measures, we consider a complex ellipticity quantity composed of two components, for example  $\epsilon = \epsilon_1 + i \epsilon_2$.  For a perfect ellipse with an axial ratio $\beta$ and orientation $\phi$, measured counter clockwise from the horizontal axis, ellipticity parameters are given by
\be
\left(
\begin{array}{c}
\epsilon_1 \\
\epsilon_2
\end{array}
\left)
= \frac{\beta-1}{\beta+1}
\right(
\begin{array}{c}
\cos 2\phi \\
\sin 2 \phi
\end{array}
\right) \, .
\label{eqn:e1e2}
\ee
As we focus this section on systematics that are related to the PSF, we construct a general model for shear measurement with a systematic error term that is linearly proportional to the PSF ellipticity, as first proposed by \citet{Baconcsys}.  We use the following model to determine the level of any residual distortions which would result from an incomplete correction for the PSF 
\be
\epsilon^{\rm obs} = \epsilon^{\rm int} + \gamma + \eta + \bm{A}_{\rm sys}^T \, \bm{e^\star}  \, .
\label{eqn:psfres_model}
\ee
Here $\epsilon^{\rm obs}$ is the observed shear estimator, $\epsilon^{\rm int}$ is the intrinsic galaxy ellipticity, $\gamma$ is the true cosmological shear that we wish to detect,  and $\eta$  is the random noise on the shear measurement whose amplitude depends on the size and shape of the galaxy in addition to the signal-to-noise of the observations.
An optimal method applied to an optimal survey will yield a random noise distribution that is significantly narrower than the intrinsic ellipticity distribution with $\sigma_\eta \ll \sigma_\epsilon^{\rm int}$ in the typical signal-to-noise regime of the data.
The systematic error term in equation~\ref{eqn:psfres_model} is given by $\bm{A}_{\rm sys}^T \, \bm{e^\star}$.  Here  \bm{e^\star} is a complex $N$ dimensional vector of PSF ellipticity at the position of the galaxy in each of the $N$ dithered exposures of the field.    In the case where the galaxy is not imaged in a particular exposure, as a result of differing chip gap and edge positions in the dithered exposures, we set the relevant exposure component of $\bm{e^\star}$ equal to zero.  
$\bm{A}_{\rm sys}$ is the amplitude of the true systematic PSF contamination which we construct as a vector of length $N$ containing the average fraction of the PSF ellipticity that is residual in the shear estimate in each individual exposure\footnote{For a single exposure image, such that $N=1$, a circular, unsheared galaxy measured with infinite signal-to-noise would have an observed ellipticity $\epsilon^{\rm obs} = A_{\rm sys} e^\star$, where the exact value of $A_{\rm sys}$ depends on how well the PSF correction has performed.  A measurement of $A_{\rm sys}$ from the data can determine what fraction of the PSF ellipticity contaminates the final shear estimate.}.   
For an unbiased shear estimate $\bm{A}_{\rm sys}^T \, \bm{e^\star} = 0$.  

We consider a series of different two point correlation functions $\xi_{\pm}$ using the shorthand notation  $\langle a b \rangle$ to indicate which two ellipticity components $a$ and $b$ are being correlated using the following data estimator: 
\be
\xi_{\pm}(\theta) = \langle \epsilon \epsilon \rangle = \frac{\sum w_i w_j \left[ \epsilon_{\rm t} (\bm{x_i}) \epsilon_{\rm t} (\bm{x_j}) \, \pm \, \epsilon_\times (\bm{x_i}) \epsilon_\times (\bm{x_j})
\right]}{
\sum w_i w_j } \, ,
\label{eqn:xipm_est}
\ee
where in this particular example we are correlating the observed galaxy ellipticities, and the weighted sum is taken over galaxy pairs with angular separation $|\bm{x_i - x_j}|=\theta$.   The tangential and cross ellipticity parameters $\epsilon_{{\rm t},\times}$ are the ellipticity parameters in equation~(\ref{eqn:e1e2}) rotated into the reference frame joining each pair of correlated objects.   In the derivation that follows we will use this shorthand notation to indicate the correlation between galaxy ellipticity, ellipticity measurement noise, PSF ellipticity and shear following the same construction of the estimator in equation~(\ref{eqn:xipm_est}).  We base our systematics analysis on this type of statistic as in the case of the two point shear correlation $\langle \gamma \gamma \rangle $ it can be directly related to the underlying matter power spectrum that we wish to probe with weak gravitational lensing,
\be
\xi_{\pm}(\theta) = \langle \gamma \gamma \rangle = \frac{1}{2\pi}\int d\ell \,\ell \,P_\kappa(\ell) \, J_{\pm}(\ell \theta) \, , 
\label{eqn:xipm}
\ee
where $J_\pm (\ell \theta)$ is the zeroth (for $\xi_+$) and fourth (for $\xi_- $) order Bessel function of the first kind and $P_\kappa(\ell)$ is the convergence power spectrum at angular wave number $\ell$ \citep[see][for more details]{Bible}.  In the case where there are no systematic errors and no intrinsic alignment of nearby galaxies, equation~(\ref{eqn:xipm_est}) is an accurate estimate of the right hand side of equation~(\ref{eqn:xipm})  \citep[see the discussion in][and references therein]{HeymansIA06, BJ11}.
 
\subsection{Cosmological Simulations: CFHTLenS clone}
\label{sec:clone}

In the analysis that follows we quantify the significance of any residual systematic error in the data by comparing our results with a cosmological simulation of CFHTLenS that we hereafter refer to as the `clone'.
The core input of the `clone' comes from 184 fully independent three-dimensional N-body numerical lensing simulations, where light cones are formed from line of sight integration through independent dark matter particle simulations, without rotation \citep{Clone2012}.  The simulated cosmology matches the 5-year {\it Wilkinson Microwave Anisotropy Probe (WMAP5)} flat $\Lambda$CDM cosmology constraints from \citet{WMAP5} and we adopt this cosmology where necessary throughout the paper, noting that our results are insensitive to the cosmological model that we choose.  Each high resolution simulation has a real space resolution of 0.2 arcmin in the shear field and spans 12.84 square degrees sampled at 26 redshift slices between $0<z<3$.  The two-point shear statistics measured in real-space from the simulations closely matches the input, dark matter only, theory from $0.5 \ls \theta \ls 40$ arcmin scales at all redshifts \citep{Clone2012}.  Being able to recover small-scale real-space resolution of the simulated shear field is crucial for our comparison analysis of systematic errors on these angular scales.

We use each independent line of sight in the simulation to create different cosmological realizations of each MegaCam field in CFHTLenS.  We ensure that the galaxy distribution, survey masks and redshifts from the data are exactly matched in the simulations, and assign galaxy shear $\gamma$ from the lensing simulations by linearly interpolating between the fine redshift slices in the simulations to get a continuous redshift distribution.  The sum of the noise $\eta$ and intrinsic ellipticity distribution $\epsilon^{\rm int}$ are assigned on a galaxy by galaxy basis by randomizing the corresponding measured galaxy orientation in the data, such that $|\epsilon^{\rm int} + \eta | = |\epsilon^{\rm obs}|$.  This step assumes that, on average, the true shear $\gamma$ contribution to the observed ellipticity $\epsilon^{\rm obs}$ is small in comparison with the measurement noise $\eta$ and intrinsic ellipticity distribution $\epsilon^{\rm int}$.   Finally, each galaxy in the `clone' is assigned a corresponding PSF ellipticity, for each exposure.  This is given by the PSF model ellipticity $\bm{e^\star}$, as measured from the data, at the location of the galaxy whose position in the MegaCam field matches the simulated galaxy position in the `clone'.

\subsection{The star-galaxy cross correlation function}
\label{sec:stargal}
In order to assess the level and significance of PSF-related systematics in the data we measure the two-point star-galaxy cross correlation function $\bm{\xi}_{\rm sg} = \langle \epsilon^{\rm obs} \bm{e^\star} \rangle$ which, using our linear shear measurement model (equation~\ref{eqn:psfres_model}), can be written as
\be
\bm{\xi}_{\rm sg} = \langle \epsilon^{\rm obs} \bm{e^\star} \rangle  =  \langle \epsilon^{\rm int} \bm{e^\star} \rangle +  \langle \gamma \, \bm{e^\star} \rangle +  \langle \eta \, \bm{e^\star} \rangle + \mathbfss{C} \bm{A}_{\rm sys} \, .
\label{eqn:sg}
\ee
$\mathbfss{C}$ here is given by the covariance matrix of PSF ellipticities between exposures such that $C_{ij} = \langle e^\star_i e^\star_j \rangle$ with $i$ and $j$ denoting the different $N$ exposures.  We assume that $\bm{A}_{\rm sys}$, the changing fraction of PSF contamination in each exposure, does not vary across the field of view\footnote{If $\bm{A}_{\rm sys}$ were dependent on position in the image or galaxy properties, for example, the method we are proposing to isolate PSF contamination would be sensitive to the average value $ \langle \bm{A}_{\rm sys} \rangle$.}. 

The derivation that follows in Section~\ref{sec:Anoise} is general for both the $\xi_{+}$ and $\xi_{-}$ components of each correlation function and applies to any angular separation probed $\theta$, and both a model and an observed stellar measurement of the PSF ellipticity.  Our primary systematics analysis however inspects only the zero-lag star-galaxy correlation $\bm{\xi}_{\rm sg}(\theta=0)$, hereafter $\bm{\xi}_{\rm sg}(0)$, using the model of the PSF ellipticity to determine $\bm{e^\star}$ at the location of each galaxy.    At zero-lag the estimator in equation~\ref{eqn:xipm_est} for the star-galaxy correlation reduces to
\be
\bm{\xi}_{{\rm sg} \pm}(0) = \frac{\sum w_i \left[ \epsilon_{1} (\bm{x}_i) \bm{e_1^\star}(\bm{x}_i) \, \pm \, \epsilon_2 (\bm{x}_i) \bm{e_2^\star} (\bm{x}_i)
\right]}{
\sum w_i} \, .
\label{eqn:xipm_zero}
\ee
The motivation for this is as follows: consider a data set where the PSF model and correction is exact such that observed galaxy ellipticities are uncorrelated with the PSF and $\bm{\xi}_{\rm sg}(0)$ is consistent with zero.   If we cross-correlate the same galaxy ellipticities with the PSF ellipticity at some distance $\theta$, $\bm{\xi}_{\rm sg}(\theta)$ will continue to be consistent with zero because the galaxies have intrinsically random orientations.   Instead, now consider a data set where there has been an error in the measurement of the PSF model or an error in the PSF model correction.  In this case the star-galaxy cross correlation at the location of the galaxies $\bm{\xi}_{\rm sg}(0)$ is now non-zero. At larger separations however, $\bm{\xi}_{\rm sg}(\theta)$  may be zero or non-zero depending on the variation of the PSF autocorrelation function.  Hence, for the detection of systematics by star-galaxy correlation, we argue that there is little information to be gained from measurements of $\bm{\xi}_{\rm sg}(\theta)$ for $\theta>0$ as it is an error in the local PSF model or local PSF correction that creates this form of systematic error.     

In the presence of systematics we would expect to detect a signal in the `+' component of the zero-lag star-galaxy cross correlation.  For systematics that are dependent on the ellipticity direction we would also expect to detect a signal in the `-' component (see equation~\ref{eqn:xipm_zero}).   The ellipticity direction dependence of any PSF residuals is, however,  expected to be weak, which we confirm in Section~\ref{sec:2pt}.  Our zero-lag systematic error analysis that follows therefore focuses on the `+' component only.  We return to the `-' component of the two-point correlation function in Section~\ref{sec:2pt}.

\begin{figure}
   \centering
   \includegraphics[width=3.5in, angle=270]{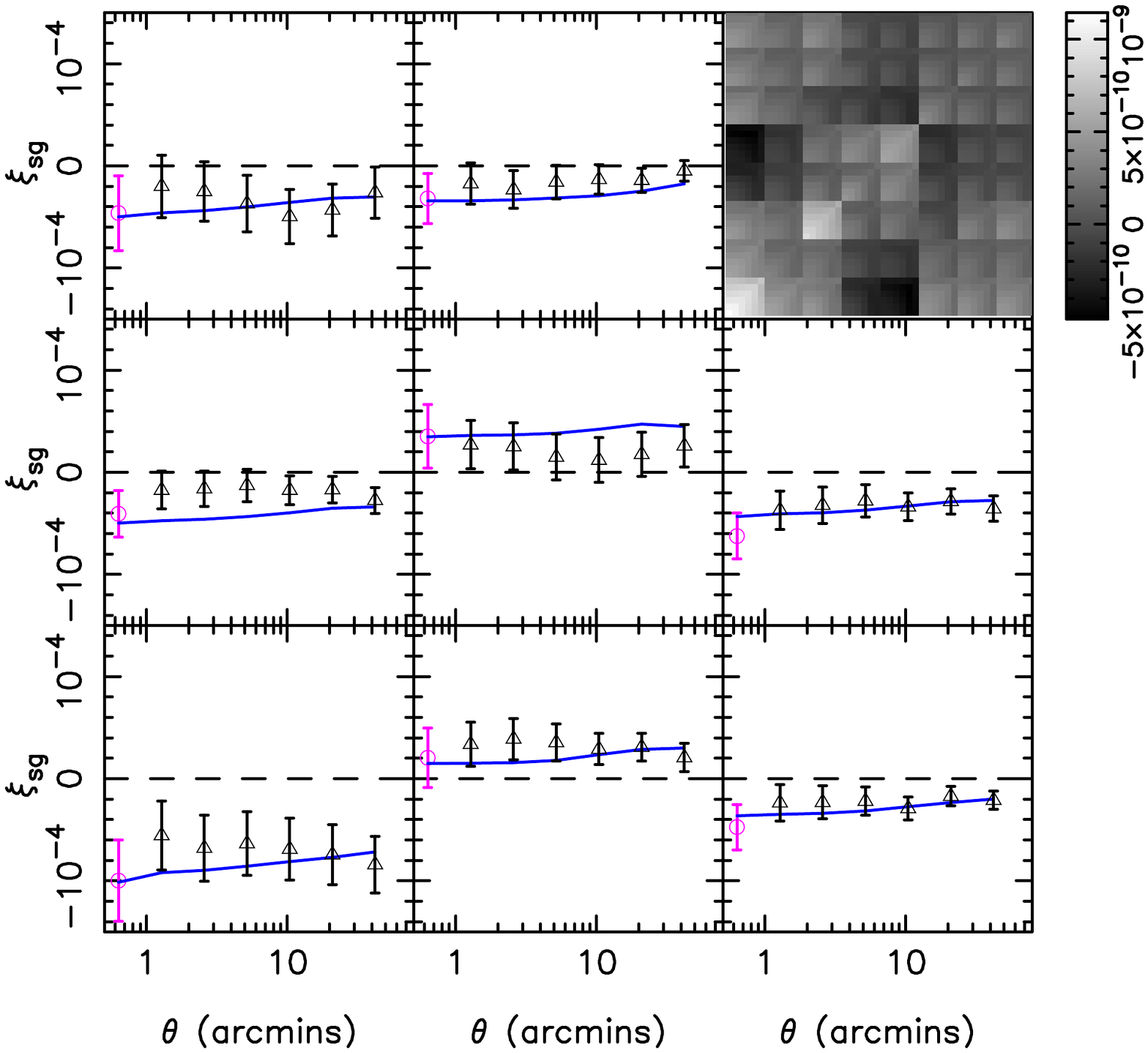} 
   \caption{The star-galaxy cross correlation function $\xi_{\rm sg}(\theta)$ for the eight individual exposures in example field W1m0m0 as a function of angular separation (triangles, where each panel is a different exposure).  The measured angular correlation function in each exposure can be compared to the predicted angular star-galaxy correlation (equation~\ref{eqn:sgtheta}, shown as a curve) calculated using only the zero separation measure $\xi_{\rm sg}(0)$ (shown offset, circle).  The correlation between the exposures and angular scales is shown in the covariance matrix of the data points in the upper right panel.  Each block shows one of the eight exposures and contains a $6 \times 6$ matrix showing the correlation between the angular scales.  The greyscale bar shows the amplitude of the values in the matrix.}
   \label{fig:example}
\end{figure}

To add further to our argument, with a measure of the zero-lag star-galaxy correlation $\bm{\xi}_{\rm sg}(0)$ we can use equation~\ref{eqn:psfres_model} to make a prediction of the star-galaxy correlation at any angular scale\footnote{Note that for a single exposure image, equation~\ref{eqn:sgtheta} reduces to $\xi_{\rm sg}(\theta_{\rm ab}) \approx 
\xi_{\rm sg}(0) \, \langle e_{\rm a}^\star e_{\rm b}^\star \rangle /  \langle e^{\star 2} \rangle$ where a and b here indicate objects separated by a distance $\theta_{\rm ab}$.  This relationship assumes that the amplitude and angular variation of the first three terms on the right hand side of equation~\ref{eqn:sg} is small in comparison to the amplitude and angular variation of the star-star auto-correlation function.}
using
\be
\bm{\xi}_{\rm sg}(\theta) \approx \mathbfss{C}^{-1}_0 \bm{\xi}_{\rm sg}(0)  \, \mathbfss{C}_\theta \, ,
\label{eqn:sgtheta}
\ee
where $\mathbfss{C}_0$ is the measured covariance matrix of PSF ellipticities between exposures at zero-lag and $\mathbfss{C}_\theta$ is the same PSF measurement but for sources at separation $\theta$.   Figure~\ref{fig:example} demonstrates this by comparing the predicted signal (equation~\ref{eqn:sgtheta} shown as a curve) with the star-galaxy cross correlation function $\xi_{\rm sg}(\theta)$ (shown as triangles) measured in the eight individual exposures in example field W1m0m0.  Seven of the exposures were imaged consecutively.  This field is typical of the sample of data that passes our systematics tests in Section~\ref{sec:fieldsel}.  Note that we use a scalar symbol here as we are referring to the measurement in each exposure rather than the vector which contains the measurement across all exposures.  The zero separation measure for each exposure $\xi_{\rm sg}(0)$ is shown offset in each panel (circle).  The correlation between the exposures and angular scales is shown in the covariance matrix in the upper panel to warn the reader that `chi-by-eye' of this data will fail.  Each block shows one of the eight exposures and contains a $6 \times 6$ matrix showing the degree of correlation between the 6 measured angular scales.  As shown in the side greyscale panel, the amplitude of the matrix is small making it sensitive to measurement errors and in order to estimate a stable covariance matrix we require a very computationally expensive bootstrap analysis of the data.  We have therefore only made a detailed comparison on 10\% of our fields, performing a $\chi^2$ goodness of fit test to the data over a range of angular scales using the model prediction from the zero-lag star-galaxy cross correlation function in equation~\ref{eqn:sgtheta}.  In all cases we find the prediction is a reasonable model for the measured star-galaxy correlation.  We also repeated the analysis for our sample fields using the measured stellar object ellipticities in contrast to the model PSF ellipticity.  Whilst our measurement errors increased, our findings were unchanged such that for the remainder of our systematics analysis we conclude that we can safely consider only the zero-lag star-galaxy cross correlation function $\bm{\xi}_{\rm sg}(0)$ as calculated using the model PSF ellipticity.   

\subsection{Estimating the level of PSF anisotropy contamination}
\label{sec:Anoise}

Assuming the linear shear measurement model of equation~\ref{eqn:psfres_model} is a good description of the systematics within the data, the systematic error contribution $\Delta\xi$ to the cosmological measure of the two-point shear correlation function  $\xi = \langle \epsilon^{\rm obs} \epsilon^{\rm obs} \rangle $ is given by
\be
\Delta\xi_{\rm sys} = \bm{A}_{\rm sys}^T  \mathbfss{C} \bm{A}_{\rm sys} \, ,
\label{eqn:dxisys}
\ee
which can be estimated from the data via\footnote{Note that for a single exposure image, equations~\ref{eqn:dxisys} and~\ref{eqn:deltaeobs} reduce to the more familiar results of \citet{Baconcsys} with $\Delta\xi_{\rm sys} = A^2  \langle e^\star e^\star \rangle$ and $\Delta \xi_{\rm obs} =  \langle \epsilon^{\rm obs} e^\star \rangle^2 /  \langle e^\star e^\star \rangle$.}
\be
\Delta \xi_{\rm obs}  = \bm{\xi}_{\rm sg}^T \mathbfss{C}^{-1} \bm{\xi}_{\rm sg} \, .
\label{eqn:deltaeobs}
\ee
When calculating $\Delta \xi_{\rm obs}$ from a very large area of data, such that the PSF is fully uncorrelated with the intrinsic ellipticity, measurement noise and cosmological shear, the first three terms in the right hand side of equation~\ref{eqn:sg} are zero and
$\Delta \xi_{\rm obs} = \Delta \xi_{\rm sys}$.   In this case the PSF correction is deemed successful when $\Delta \xi_{\rm obs}$ is found to be consistent with zero.    This method for data verification has been applied to many previous weak lensing surveys \citep[see for example][]{Baconcsys} but only in an ensemble average across the full survey area and for single stacked images.  By taking an ensemble average of $\Delta \xi_{\rm obs}$ across the survey, one explicitly assumes that the true level of PSF contamination that we wish to estimate is independent of the variations in the quality of the data.  For ground-based observations where the data quality varies considerably we might expect our ability to remove the PSF to be reduced in some particular instances, for example poorer seeing or low signal-to-noise data.  By determining $\Delta \xi_{\rm obs}$ averaged across the survey we could easily miss a small fraction of the data which exhibit a strong PSF residual.  In the worst case scenario, as the CFHTLS PSF exhibits strong variation in direction and amplitude between exposures, PSF residual effects could easily cancel out in an ensemble average (see Section~\ref{sec:csysB} for further discussion on this point).  We therefore choose to apply this methodology to individual one square degree MegaCam fields (hereafter referred to as a field), in order to identify fields with exposures that exhibit a strong PSF residual.

\begin{figure}
   \centering
   \includegraphics[width=2.8in, angle=270]{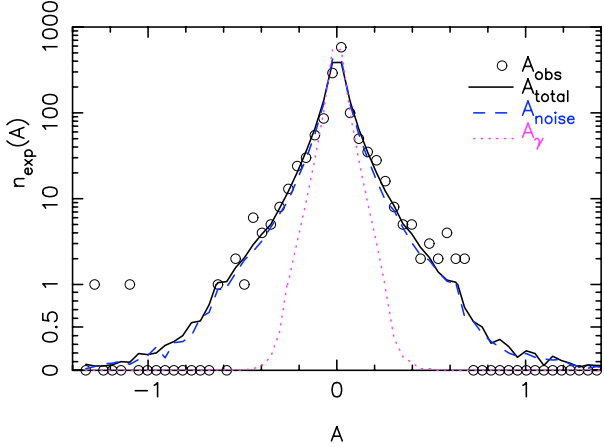} 
   \caption{ The distribution of the components of $A_{\rm obs}$ for individual exposures in CFHTLenS.  The open symbols show the distribution for the full data set.  As the number of exposures is discrete we change from a log to linear scale below $n(A) = 1$.  The data can be compared with the different components of the star-galaxy cross correlation function $\bm{\xi}_{\rm sg}$ as measured from the simulated CFHTLenS `clone' data.  The dashed curve shows the contribution to the star-galaxy cross correlation function from the intrinsic ellipticity distribution $A_{\rm noise}$ and the dotted curve shows the contribution from chance alignments with the galaxy shear field $A_\gamma$,  demonstrating that significant star-galaxy correlations can be measured from chance alignments of the PSF with the galaxy shear and intrinsic ellipticity field in one-square degree regions.}
   \label{fig:Acomp}
\end{figure}

For the individual analysis of a one square degree field, we can no longer assume that $\Delta \xi_{\rm obs} = \Delta \xi_{\rm sys}$ as the three noise terms in the right hand side of equation~\ref{eqn:sg} can be significant simply from a chance alignment of cosmological shear, random measurement noise or intrinsic ellipticities with the PSF.  Using one square degree patches of the CFHTLenS `clone' (see Section~\ref{sec:clone}) we find $\Delta \xi_{\rm obs} > \Delta \xi_{\rm sys}$ even when $\bm{A}_{\rm sys} = 0$.     
To illustrate this point we multiply each component in equation~\ref{eqn:sg} by the inverse PSF covariance $\mathbfss{C}^{-1}$ to define $\bm{A}_{\rm obs}$, 
\be
\bm{A}_{\rm obs} = \mathbfss{C}^{-1} \bm{\xi}_{\rm sg} = \bm{A}_{\rm noise}+  \bm{A}_{\rm \gamma} + \bm{A}_{\rm sys}  \, ,
\ee
such that $\bm{A}_{\rm obs}$ would be equal to $\bm{A}_{\rm sys}$, the scale of the true residual PSF signal in each exposure, if the noise terms $\bm{A}_{\rm noise}$ and $\bm{A}_{\gamma}$ could be ignored, where
\be
\bm{A}_{\rm noise} = \mathbfss{C}^{-1} \langle (\epsilon^{\rm int} + \eta) \, \bm{e^\star} \rangle \, ,
\ee
\be
\bm{A}_{\gamma} = \mathbfss{C}^{-1} \langle \gamma \,\bm{e^\star} \rangle \, .
\ee
For each CFHTLenS field we first calculate $ \mathbfss{C}^{-1}$ from the measured PSF model in each exposure.  We then calculate the distribution of values we measure for $\bm{A}_{\rm noise}$ and $\bm{A}_{\gamma}$ for each field, keeping $\mathbfss{C}$ fixed, but varying $\epsilon^{\rm int} + \eta$ and $\gamma$ using all 184 independent simulations from the `clone'.  Figure~\ref{fig:Acomp} compares the distribution of values measured for each component of $\bm{A}_{\rm noise}$  (dashed) and $\bm{A}_{\gamma}$ (dotted) for all simulated realizations of the fields, normalized to the total number of exposures in the survey.  This can be compared to the total discrete number of exposures with $A_{\rm obs}$ as measured from the complete CFHTLenS data set (circles).  Note that we use a scalar symbol here as we show the distribution of measurements over all exposures in the survey rather than the vector which contains the measurement across all exposures in a particular field. 
This figure shows that the combined distribution of  $\bm{A}_{\rm noise}$ and $\bm{A}_{\gamma}$ (solid) as measured from the simulated data is generally consistent with the observed distribution of $\bm{A}_{\rm obs}$ over all CFHTLenS MegaCam imaging.  We do, however, observe some outliers from the expected distribution and indications of an increased width of the observed distribution from the simulated distribution.  This comparison reveals the presence of a systematic PSF residual signal in a small fraction of our data.  

Before we further develop our method to identify problematic data in Section~\ref{sec:pU}, we should pause to note a general cause of concern over our choice to perform this analysis on one square degree fields.  This illustration particularly highlights the width of the $\bm{A}_{\rm noise}$ distribution compared to the low levels of $\bm{A}_{\rm sys}$ that can be tolerated in a cosmological analysis.  
In order for a tomographic cosmic shear analysis of CFHTLenS not to be dominated by systematic errors we can estimate a ballpark tolerance for the level of PSF related systematics in the data \citep{AmaraRef08}.  Assuming a conservative average PSF ellipticity of $e^\star \sim 0.1$, we require $\bm{A}_{\rm sys} \ls 0.01$ for CFHTLenS.  This is an order of magnitude less than the width of the distribution of $\bm{A}_{\rm noise}$ on a field by field basis.  This analysis will therefore only be able to identify the most problematic data in our sample, but we show how crucial this step is in Section~\ref{sec:2pt}.  Furthermore we note that for the first weak lensing surveys imaged with quarter square degree field of view cameras and less, this type of field by field analysis would have been even less sensitive to PSF errors. For future space-based surveys with a stable PSF pattern between exposures and fixed data quality, data could be grouped, however, by PSF pattern to allow for the extension of this analysis over much larger areas.  Such grouping would reduce the width of the $\bm{A}_{\rm noise}$ and $\bm{A}_{\gamma}$ distributions, allowing for the detection of even the lowest level of PSF residuals.  It should be noted however that as the $\bm{A}_{\rm noise}$ distribution decreases, this test could become sensitive to the fiducial cosmology used in the simulations to compute $\bm{A}_{\gamma}$.  In addition, in a low noise regime, very aggressive cuts on $\bm{A}_{\rm obs}$ could potentially risk the rejection of high shear regions of the data.  As these high shear regions have a true value of $\bm{A}_{\gamma}$ that live in the wings of the distribution estimated from the simulations, any cuts must avoid misinterpreting high shear regions as data with a high $\bm{A}_{\rm sys}$ instead.  As Figure~\ref{fig:Acomp} shows that the $\bm{A}_{\rm noise}$ distribution is significantly larger than $\bm{A}_{\gamma}$ we clarify that we are not in this regime for our analysis of CFHTLenS.  

Finally we remind the reader that using the cross-correlation between the galaxy and the PSF model ellipticities does allow us to test for systematic errors in the data, but does not allow us to isolate the cause of such errors if they are detected.  The most likely cause would result from residual errors in modeling the PSF.  In addition, we also expect noise biases to introduce some low-level of cross-correlation as described in Appendix C of \citet{Miller2012}.  Using a simplified example of Gaussian galaxy and PSF profiles,  \citet{Miller2012} show that the mean of the resulting ellipticity likelihood surface is biased away from the true ellipticity value, usually leading to a negative multiplicative calibration bias, discussed in Section~\ref{sec:empcor} \citep[see also][]{RefKac,Kacprzak}.    The inclusion of a PSF ellipticity in this Gaussian analysis revealed that the mean of the ellipticity likelihood surface is also biased towards the PSF ellipticity.  The extent of this effect depends on the PSF size and ellipticity and the galaxy size, ellipticity and signal-to-noise ratio.  It is argued that the bias is also likely to depend on the radial surface brightness profiles of the PSF and galaxy \citep{Miller2012}.  At a low signal-to-noise ratio,
a star-galaxy cross-correlation term may therefore develop, although its amplitude is hard to predict.  We do not, in this paper, attempt to separate the effect of this bias from PSF modeling errors, but instead treat any measured net effect as being indicative of an overall bias in the data, regardless of its origin.

\subsection{Identification of observations with significant residual PSF errors}
\label{sec:pU}
In order to identify observations with significant residual PSF errors we wish to calculate the probability that the measured systematic contribution to the two-point cosmic shear correlation function $\Delta \xi_{\rm obs}$ is consistent with $\bm{A}_{\rm sys} = 0$.  To do this we are required to take into account noise contributions from the intrinsic ellipticity, random measurement noise and shear as illustrated in Figure~\ref{fig:Acomp}.

We first define a vector $\bm{U}$, whose components are uncorrelated, such that  $\Delta \xi_{\rm obs} = \bm{U}^T\bm{U}$  with
\be
\bm{U} = \mathbfss{L}^{-1/2} \mathbfss{V} \bm{\xi}_{\rm sg} \, ,
\label{eqn:U}
\ee
where we have decomposed $\mathbfss{C}$ into independent eigenvalues and eigenvectors such that $\mathbfss{C} = \mathbfss{V}^T\mathbfss{LV}$ \citep[see singular value decomposition section of][for more details]{Numrec}.    
There are two noise contributions to $\bm{U}$.  The first comes from the sum of the intrinsic ellipticity, measurement noise and shear contributions to $\bm{\xi}_{\rm sg}$ that we wish to capture in our analysis.  This we estimate by measuring the variance of $\bm{U}$ over 184 `clone' realizations of $\bm{\xi}_{\rm sg}$ where $\mathbfss{C}$ is measured from the data.  To do this we assume the null hypothesis that $\bm{A}_{\rm sys} = 0$ which is reasonable as, in the case where the systematics are non-zero, this method would underestimate the error and make the systematics appear more significant.     
The second noise contribution arises from any measurement error in $\mathbfss{C}$, which, in its inversion in equation~\ref{eqn:deltaeobs}, could lead to instabilities in the calculation of $\Delta \xi_{\rm obs}$.  For CFHTLenS the PSF correlation between exposures, given by the off-diagonal components of $\mathbfss{C}$, are typically very small, $\sim 10^{-10}$, and hence subject to measurement error.  Considering that the PSF ellipticities are typically of the order $10^{-2}$ these off-diagonal terms reveal how uncorrelated and hence how unstable the PSF pattern is between even consecutive exposures.   We estimate a measurement error on $\mathbfss{C}$ using a bootstrap technique dividing the model PSF measured at each galaxy location, $\bm{e^\star}$, into CCD chip-sized groups based on their location in the MegaCam field of view.  We calculate the variance introduced to each component of $\bm{U}$ measured from the data when $\mathbfss{C}$ is estimated from different CCD chips in the field of view selected at random (with repetition).  We find the PSF covariance measurement error term to be sub-dominant for the majority of components of $\bm{U}$.     For the components of $\bm{U}$ with eigenvalues less than a few percent of the maximum eigenvalue however, we find that the measurement errors on $\mathbfss{C}$ start to dominate, leading to unstable measures of $\Delta \xi_{\rm obs}$.  

In order to ensure our analysis is not dominated by uncertainty in our measurement of $\mathbfss{C}$,  we choose to only consider the $M$ eigenmodes of $\mathbfss{C}$ which contribute less than 20\% of the total noise $\sigma_i$ as measured on each component $U_i$.  The total noise $ \sigma_i$ per component combines in quadrature the variance of $U_i$ over the `clone' simulations and the variance on $U_i$ between different bootstrap PSF realizations.  To be clear, it is this latter component which needs to contribute less than 20\% of the total noise for the eigenmode to be considered in the analysis.
The discarded modes of the PSF covariance $\mathbfss{C}$ typically have eigenvalues less than $\sim 5\%$ of the maximum eigenvalue.  This step can therefore be considered as a principal component decomposition of the PSF covariance matrix $\mathbfss{C}$ to use $\sim 95\%$ of the information within the covariance matrix to calculate $\Delta \xi_{\rm obs}$,  removing the other $\sim 5$\% of the information in the PSF covariance matrix that is attributed to measurement noise.  The exact value of the eigenvalue cut is, however, motivated on a field by field basis, based on the measured noise in $\mathbfss{C}$.  As any systematics are likely to be dependent on the main PSF correlation between fields, the application of this `de-noising' method to the PSF covariance matrix is not expected to bias our results.

Once the number of $M$ reliable eigenmodes of the PSF covariance matrix has been established, we can measure a probability that the field has $\Delta \xi_{\rm obs}$ that is consistent with $\bm{A}_{\rm sys} = 0$ by calculating 
\be
\chi^2 = \sum_{i=1}^{M}  \frac{U_i^2}{\sigma_i^2}
\label{eqn:chisq}
\ee
where $M$ is the number of eigenmodes remaining in the analysis and the total error $\sigma_i$ on each component $U_i$ combines the two errors described above in quadrature.  From equation~\ref{eqn:chisq} we then determine a probability that $\Delta \xi_{\rm obs}$ is consistent with zero systematics; $p(\bm{U}=0) = p(\chi^2 | \nu)$.  This is calculated from an incomplete gamma function \citep{Numrec} with the number of degrees of freedom $\nu$ given by the number of eigenmodes in the analysis $M$.  For the standard $N=7$ exposure field we typically find $M=4$.

\section{Results and Analysis}
\label{sec:res}

In Section~\ref{sec:method} we presented a general method to utilize the star-galaxy cross correlation function $\bm{\xi}_{\rm sg}$ as measured in each exposure, to isolate data where residual signals remain that are correlated with the PSF.   In this section we discuss the necessary calibration corrections that we apply to data and the application of our star-galaxy cross correlation analysis to select a sample of fields that have a star-galaxy cross correlation signal that is consistent with noise.  We then investigate which fields fail our systematics tests.

\subsection{Calibration Corrections}
\label{sec:empcor}

Calibration corrections are a standard feature of weak lensing analyses.  They typically consist of a multiplicative component $m$, calibrated through the analysis of simulated images, and an additive component $c$, calibrated empirically from the data, 
\be
\epsilon^{\rm obs} = (1 + m) [\gamma +\epsilon^{\rm int}] + c \, ,
\label{eqn:stepmc}
\ee
where $\epsilon^{\rm obs}$ is the observed shear estimator for the measurement method.  This shear estimator is typically an estimate of galaxy ellipticity, which {\em lens}fit defines as in equation~\ref{eqn:e1e2}, but differing parameterization between methods is common. 
With a perfectly calibrated analysis pipeline, the resulting shear estimator is unbiased with $m = c = 0$.  This is a level of accuracy which can never be confirmed, owing to noise in the simulated data, but it is a level of accuracy which is aspired to and has not yet been achieved to better than percent level precision.  Many shear measurement methods have been calibrated in the recent GREAT10 challenge which confirmed previous image analysis competitions; all tested shape measurement methods suffer from noise bias with $|m|$ increasing as the signal-to-noise of the galaxy decreases \citep{GREAT10}.  Whilst $c$ is often found to be negligible in image simulations, this is not always the case with data, as shown by the two most recent cosmic shear analyses by \citet{TS10} and \citet{Huff}. \citet{TS10} present a cosmic shear analysis of the Hubble Space Telescope (HST) COSMOS survey.  They use image simulations from \citet{STEP2} to calculate $m = -0.078 (S/2)^{-0.38}$ where $S$ is a scaled measure that is proportional to the signal-to-noise of the galaxy whose shear measurement is to be calibrated. $S=2$ corresponds to a signal-to-noise ratio of $\nu_{\rm SN}\sim 7$ \citep{erben}.  An additive calibration $c$ is also calculated empirically from the data as a function of flux, observation date, sky noise, position on the image and galaxy size in order to account for the effects of charge transfer inefficiency (CTI) in the Advanced Camera for Surveys on HST.  This space-based empirical correction can be as large as $c_1 \sim 0.04$ \citep{RhodesPSF} and affects both components of the ellipticity \citep{TS10}.  \citet{Huff} present a cosmic shear analysis of the deep Stripe 82 data from SDSS.  They simulate SDSS Stripe 82 depth images using a public software package called SHERA \citep{SHERA} and calculate an average calibration correction that combines
both a noise bias term and a responsivity correction.  The responsivity correction is related to the width of the intrinsic ellipticity distribution and is expected to result in a correction in the region of 0.8 to 0.9 in terms of $m$, where the exact value is dependent on the galaxy population.  The SHERA simulation calibrated combined noise bias and responsivity correction corresponds to an average correction of $m=0.776$ applied to each galaxy.  An average additive calibration to the $\epsilon_1$ component of the ellipticity of amplitude $c_1 = -0.002$ is also applied to each galaxy to account for a bias which is believed to be introduced by the preferential direction of the elongated SDSS photometry masks.  

\citet{Miller2012} detail the CFHTLenS image simulations used to quantify the required calibration for {\em lens}fit finding an additive correction $c$ consistent with zero for the simulated data.  The multiplicative calibration term $m$ was, however, found to be significant and dependent on both galaxy signal-to-noise $\nu_{\rm SN}$ and size $r$ with
\be
m(\nu_{\rm SN},r) = \frac{\beta}{\log(\nu_{\rm SN})} \exp^{-r \, \alpha \, \nu_{\rm SN} } \, ,
\label{eqn:alphabeta}
\ee
and a best-fit $\alpha = 0.057$ and $\beta=-0.37$.   On a weighted average, this corresponds to a 6\% correction with $\langle m \rangle = 0.94$.   As discussed in \citet{Miller2012} an unbiased way to apply this calibration correction is through a weighted ensemble average correction, rather than dividing by $(1+m)$ on a galaxy-by-galaxy basis.   In the calculation of our systematics test parameter $\bm{U}$ (equation~\ref{eqn:U}) we therefore calculate the calibration correction $\bm{U}^m$ given by 
\be
\bm{U}^m = \mathbfss{L}^{-1/2} \mathbfss{V} \langle \, [1+m(\nu_{\rm SN},r) ] \, \bm{e^\star} \rangle \, ,
\label{eqn:Um}
\ee
where the weighted average is taken over the same set of galaxies used to calculate $\bm{U}$.  We remind the reader of the shorthand notation used in this paper, as described in equation~\ref{eqn:xipm_est}.  The components of the calibrated systematics test parameter we use in our $\chi^2$ analysis (equation~\ref{eqn:chisq}) are then given by $U_i^{\rm cal} = U_i / U_i^{\rm m}$.  

CFHTLenS image simulations are also used to investigate intrinsic ellipticity bias.   \citet{Miller2012} find evidence that the {\em lens}fit weights are very weakly biased in the sense that the galaxies that are intrinsically oriented perpendicular to the PSF have slightly narrower likelihood surfaces in comparison to those galaxies oriented parallel to the PSF.  This would be an unwanted side-effect of any model fitting code that does not attempt to correct for this effect, and indeed is something which is also seen in non-model fitting shape measurement methods \citep{STEP1}.  If significant for CFHTLenS, this would be identified with our star-galaxy cross correlation analysis and the problematic data would be removed accordingly.    

\begin{figure}
   \includegraphics[width=3.7in, angle=0]{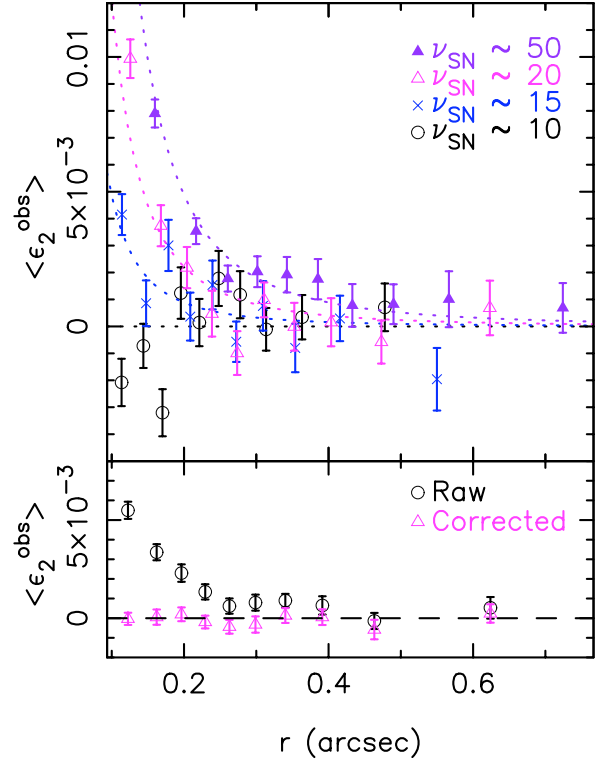} 
   \caption{Additive Calibration Correction.  The upper panel shows the weighted average $\langle \epsilon_2^{\rm obs} \rangle$ as a function of galaxy size, $r$, for four galaxy samples split by signal-to-noise ratio (circles, $\nu_{\rm SN} \sim 10$, crosses, $\nu_{SN} \sim 15$, open triangles, $\nu_{\rm SN} \sim 20$, and filled triangles $\nu_{SN} \sim 50$ ).  It is therefore the small bright galaxies that contribute the most to the average raw measure of $\langle \epsilon_2^{\rm obs} \rangle$ shown in the lower panel (circles).  The $r$ and  $\nu_{\rm SN}$ dependent model in equation~\ref{eqn:cmodel} is fit to the data in fine bins of size and signal-to-noise ratio.  The best-fit calibration correction for each $\nu_{\rm SN}$ in the upper panel is shown by the dashed curves.  The average $\langle \epsilon_2^{\rm obs} \rangle$ when measured from the calibrated galaxy catalogues is shown to be consistent with zero (triangles, lower panel).  All errors come from a bootstrap analysis of the catalogue.}
   \label{fig:empcor}
\end{figure}

We now turn our focus to calibrating any required CFHTLenS additive calibration correction $c$ empirically from the data, even though this was found to be consistent with zero in the image simulation analysis.  The comparison of equation~\ref{eqn:stepmc} with equation~\ref{eqn:psfres_model} makes it tempting to equate $c$ with the systematic error term discussed in Section~\ref{sec:method} where $c = \bm{A}_{\rm sys}^T \, \bm{e^\star}$.  As we can remove any data with significant PSF residuals, however, such that  $\bm{A}_{\rm sys}$ is consistent with zero, any strong additive calibration term $c$ remaining in the data that passes our systematics tests must derive from an alternative source of error.  For the data that passes our systematics tests, as discussed further in Section~\ref{sec:fieldsel}, we measure an average $ \langle c_1 \rangle = \langle w \epsilon_1^{\rm obs} \rangle = 0.0001 \pm 0.0001$ that is consistent with zero. For the second component of the ellipticity however we find an additive calibration that is significantly non-zero with $ \langle c_2 \rangle = \langle w\epsilon_2^{\rm obs} \rangle = 0.002 \pm 0.0001$.  Detailed investigation showed this effect to be independent of PSF size, PSF ellipticity and galaxy type. There is however a clear dependence on galaxy size $r$, and signal-to-noise ratio $\nu_{\rm SN}$, where both parameters are determined by {\em lens}fit.  We find that it is the smallest brightest objects contributing the strongest signal to the additive calibration.  This can be seen in the upper panel of Figure~\ref{fig:empcor} which shows the weighted average $\langle \epsilon_2^{\rm obs} \rangle$ as a function of galaxy size for four galaxy samples split by signal-to-noise ratio $\nu_{\rm SN}$.  The average is taken over all fields that pass our systematics test, where the field selection is presented in Section~\ref{sec:fieldsel}. 

To quantify the additive component of the calibration correction, we model $c_2$ as a function of size and signal-to-noise as
\be
c_2 = \rm Max \left [\frac{{\it F} \log_{10}(\nu_{\rm SN}) - {\it G}}{1 + \left(\frac{r}{r_0}\right)^{\it H}}\,\, ,\,\, 0 \right] \, ,
\label{eqn:cmodel}
\ee
and use a maximum likelihood method to fit the free parameters $[{\it F,G,H},r_0]$ to the data finely binned with 20 bins in $r$ and 8 bins in $\nu_{\rm SN}$.  These were chosen such that there was sufficient resolution in size to map the upturn for small galaxy sizes and sufficient signal in the data to get a reliable fit.  The parameters are strongly degenerate with a best fit reduced $\chi^2 = 1.4$ with $[{\it F}=11.910, {\it G}=12.715, {\it H} = 2.458, r_0=0.01$"$]$.  To illustrate the calibration correction we show the best-fit model on the upper panel of Figure~\ref{fig:empcor} for the mean signal-to-noise ratio of each sample, emphasizing that the data shown here is for four broad bins in $\nu_{\rm SN}$ and that the model was fit to finer bins.  The lower panel shows the effect of the $c_2$ calibration correction when applied to each galaxy, reducing the average additive signal (shown circles) to be consistent with zero at all scales (shown triangles).   We do not calculate errors on the free parameters in the model for the additive calibration correction as, if uncalibrated, a systematic additive error would enter into the cosmic shear signal only at the level of $10^{-6}$.  Hence the error on the additive calibration correction is negligible in our error budget.  For simplicity we apply this small additive $c_2$ correction on a galaxy-by-galaxy basis.  This is in contrast to the ensemble average multiplicative $m$ correction (see for example equation~\ref{eqn:Um}), as unlike the division by $1+m$ on a galaxy-by-galaxy basis, the subtraction of $c_2$ is stable.

It is unclear what causes this bias in the measured $\epsilon_2$ component for the small, bright galaxies.  We can rule out the CTI that impacted \citet{TS10} as CFHTLenS has a high sky background and CTI affects predominantly the $\epsilon_1$ component.  We can also rule out the hypothesis of masking bias in \citet{Huff} as the smaller masks in CFHTLenS are generally circular or random in orientation with the exception of the stellar diffraction spike masks which, according to \citet{Huff}, would affect the $\epsilon_1$ component not $\epsilon_2$.   One immediate potential source of error that would affect $\epsilon_2$ more strongly than $\epsilon_1$ is the undersampling of the PSF as the size of the pixel in the $\epsilon_2$ direction is longer than in the $\epsilon_1$ direction.  This effect is, however, not seen in the analysis of image simulations which include the correct MegaCam pixel scale and typical CFHTLS PSFs, modeled using the best-fitting \citet{Moffat} light profile to each exposure in the survey.  As the effect is uncorrelated with PSF properties, we are therefore led to conclude that it most likely arises from the data analysis and could possibly be linked to an unknown effect in the data processing or within the CFHT MegaCam camera.   Possible examples could be a non-isotropic charge diffusion within the CCD,   non-uniform intra-pixel response, residual errors in the cosmic ray masking which have a non-isotropic behaviour or low-level long-term persistence or ghosting effects, as the CFHTLS dither strategy is in the positive $(\epsilon_1,\epsilon_2)$ direction.     We note that there is concern that a random camera or data processing related affect could also impact the PSF modeling, as the stellar objects are small and bright.  The unknown cause of this bias is therefore also potentially linked to the fraction of fields that we reject using our star-galaxy correlation systematics test.

\subsection{Field Selection}
\label{sec:fieldsel}

\begin{figure}
   \centering
   \includegraphics[width=3.8in, angle=0]{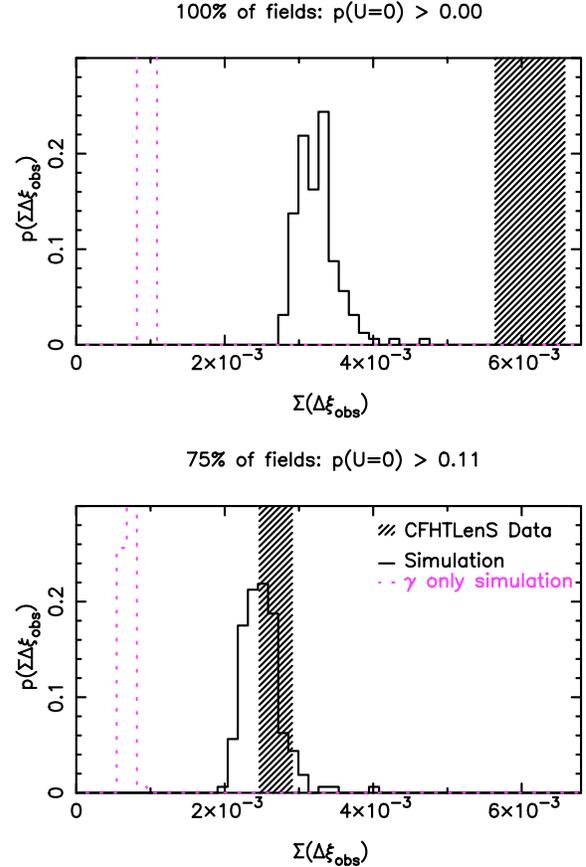} 
   \caption{Comparison of the measured $\Sigma(\Delta \xi_{\rm obs})$ (hatched) where the sum is taken over all fields (upper panel) or over the fields with a measured probability of zero systematics $p(\bm{U}=0)>0.11$ (lower panel).  These measures can be compared with the probability distribution of measuring  $\Sigma(\Delta \xi_{\rm obs})$ from the same number of fields realized in the systematics free CFHTLenS `clone' (solid).  For the full data set (upper panel), we find that the measured 
   $\Sigma(\Delta \xi_{\rm obs})$ far exceeds what is expected from the simulations.  Once a conservative cut is applied to the data (lower panel) removing 25\% of the data, we find the measured  $\Sigma(\Delta \xi_{\rm obs})$ is fully consistent with the expected distribution for the same number of simulated fields.  For comparison we also show the probability distribution of $\Sigma (\Delta \xi_{\rm obs})$ as measured from a random correlation between the pure cosmic shear $\gamma$ and the range of CFHTLenS PSFs (dashed). }
   \label{fig:pcut}
\end{figure}

For each field we calculate the systematics test parameter $\bm{U}$ (equation~\ref{eqn:U}) applying the calibration corrections described in Section~\ref{sec:empcor}.  We then calculate the probability that $\Delta \xi_{\rm obs}$ is consistent with zero systematics; $p(\bm{U}=0)$ as detailed in Section~\ref{sec:pU} and set an acceptance threshold on this probability using a method that we demonstrate in Figure~\ref{fig:pcut}.   Here, in the upper panel, we show the measured systematic error observable $\Sigma (\Delta \xi_{\rm obs})$ where the sum is taken over all fields (hatched area includes the $1\sigma$ bootstrap error on the measure).  This can be compared with the distribution of values obtained from all the different realizations of the CFHTLenS `clone' (solid).  The `clone' distribution shows the probability of measuring $\Sigma (\Delta \xi_{\rm obs})$ from the full survey area if there were no PSF residuals in the data.   Note that by definition $\Delta \xi_{\rm obs}$ at zero-lag is a positive quantity (see equation~\ref{eqn:deltaeobs}) so even for the simulated `clone' catalogues which have zero systematics, by definition, $\Sigma (\Delta \xi_{\rm obs})$ is non-zero.   For comparison we also show the distribution of $\Sigma (\Delta \xi_{\rm obs})$ that would be measured simply from a random correlation between the pure cosmic shear $\gamma$ and the range of CFHTLenS PSFs (dashed).  The significance of this signal re-iterates the points made in Section~\ref{sec:Anoise} of how important it is to take into account both the random intrinsic ellipticity noise and underlying cosmic shear in this type of systematics analysis.

The conclusion we can draw from the upper panel of Figure~\ref{fig:pcut} is that when we consider the full data set, the sum of the measured star-galaxy cross correlation is very significant compared to the expectation from the simulated `clone' catalogues.    
We therefore set a criterium that selects only those fields above a tunable threshold probability that $\Delta \xi_{\rm obs}$
is consistent with zero systematics; $p(\bm{U}=0)$.  By increasing the cut on $p(\bm{U}=0)$ the measured systematic error observable $\Sigma (\Delta \xi_{\rm obs})$ decreases rapidly as using $p(\bm{U}=0)$ for our selection criteria preferentially rejects the fields with the strongest systematic residual errors.   As the number of fields in the analysis decreases, the $\Sigma (\Delta \xi_{\rm obs})$  expected from the `clone' also decreases.  This is because it is summed only over the number of fields remaining in the analysis and there are fewer positive numbers to sum.   We continue this rejection process until the $1\sigma$ confidence region on our measured systematic error observable $\Sigma (\Delta \xi_{\rm obs})$ is in agreement with the peak of the probability distribution expected for this quantity from the same number of fields in the `clone' simulations (lower panel).   It is interesting to note that the variance of the simulated distributions also becomes consistent with the $1\sigma$ error on the measured $\Sigma (\Delta \xi_{\rm obs})$ when the threshold selection is optimised in this way.  This process sets a threshold of $p(\bm{U}=0) > 0.11$ below which we label the field as `failed'.  This leaves us with 75\% of CFHTLenS fields which pass the systematics test.  We investigate the impact of this cut for two-point cosmic shear statistics in Section~\ref{sec:2pt}. 

For a complete and detailed account of the analysis we should clarify at this point that the field selection and empirical $c_2$ additive calibration correction described here and in Section~\ref{sec:empcor} is actually calculated using a two-step iteration.  We first select fields applying only the multiplicative $m$ calibration correction (equation~\ref{eqn:Um}) as calculated from our simulated image analysis in \citet{Miller2012}.  This first-pass field selection safeguards that the empirical $c_2$ calibration correction we calculate from the selected data is unrelated to the PSF.
The additive correction that is empirically calculated from these selected fields is then applied to the full survey.  We then re-run our systematics analysis on the full survey to re-select fields which pass the systematics tests when both the multiplicative and first-pass additive calibration corrections are included.  This safeguards that in the first-pass iteration, the additive error term, now corrected by the $c_2$ calibration, did not mask the presence of PSF residuals, or appear as a PSF residual in exposures where the PSF is predominantly in the $e_2^\star$ direction.  At this second-pass iteration we lose two fields and gain seven fields into our selected clean data sample.  Finally we empirically re-calculate the additive calibration correction $c_2$ for this final set of selected fields to improve the accuracy of the correction on the final field sample.  This re-calculation introduces a small percent level adjustment to the first-pass measure and is the $c_2$ calibration that is presented in equation~\ref{eqn:cmodel}.  

Finally we discuss duplicate fields, originally imaged with an $i'$.MP9701 filter, and re-imaged,
after this initial filter was damaged in October 2007,  with the replacement $i'$.MP9702 filter.   In general, we do not distinguish between these two periods of $i'$ imaging, although the different filter response curves are of course accounted for in our photometric redshift analysis \citep{HH12}.  For the purposes of this discussion, however, we will refer to these two filters as $i'_1$ and $i'_2$.  Duplicate fields were re-imaged in order to calibrate and assess the impact of the change of filter mid-survey, in addition to some cases where preliminary concerns about the PSF in the original observations led to their re-observation.  In total 18 fields were imaged to close to full depth in both the $i'_1$ and $i'_2$ band in the optimal seeing conditions required for the lensing analysis.  In principle, we may expect the second pass $i'_2$ observation to be the best data as the re-imaging for some of these fields was undertaken to improve the PSF.   It is therefore interesting to compare the systematics results for these 18 fields, and clarify which of the fields we choose to pass in the cases where both the $i'_1$ and $i'_2$ band images pass our systematics analysis.
Out of the duplicated fields, 2 failed the systematics test in both filters, 5 passed only in $i'_1$ and 1 passed only in $i'_2$.   Out of the 10 fields which passed in both $i'_1$ and $i'_2$, we choose to select data for our scientific analysis from the filter with the highest probability $p(\bm{U}=0)$ with 4 fields selected in $i'_1$ and the other 6 in $i'_2$.  With such a small number of fields we cannot draw any significant conclusions on the dependence of systematics on the filter, or whether the preliminary concerns about the PSF in some of the duplicate observations were warranted.  

\subsection{Comparison of fields which pass and fail systematics tests}
\label{sec:scatter}
The star-galaxy cross correlation field selection presented in Section~\ref{sec:fieldsel} identifies 25\% of fields which have a significant PSF residual signal.  Whilst it is disappointing to reject this fraction of data from our cosmological analysis of CFHTLenS, it is a major step forward to be able to robustly identify the data that adds systematic error into our analysis.  Crucially we can do this in an analysis that is insensitive to assumptions about the cosmology of the Universe that we wish to determine.   Factoring in the 19\% area that is lost to masked defects such as stellar diffraction spikes (see section ~\ref{sec:masks}), our final total rejected area is 39\% of the survey.  We can compare the amount of rejected CFHTLenS data to the fraction of data rejected in previous CFHTLS analyses which totaled 30\% in \citet{HH06} and 40\% in \citet{Fu08}, which arose in both cases from a combination of masked defects and the conservative chip boundary masks that we do not apply in our analysis.

\begin{figure}
   \centering
   \includegraphics[width=3.6in, angle=0]{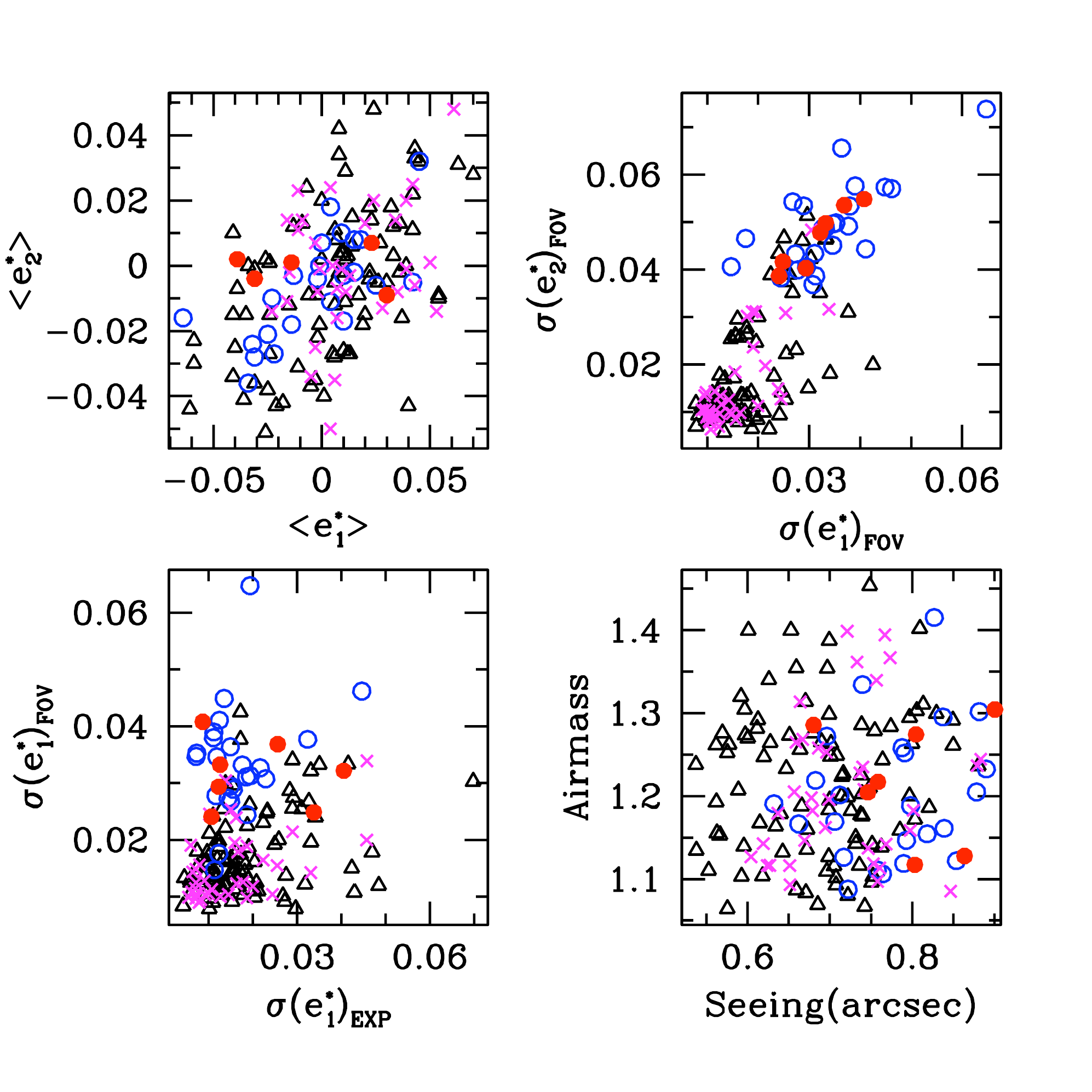} 
   \caption{Comparison of fields which pass and fail the PSF systematics test to different observables.  The open symbols indicate fields that pass.  The crosses and filled circles indicate fields that fail.  These two sets are split into data observed before the CFHT `lens flip' (open and filled circles) and the majority of the data observed after the CFHT `lens flip' (open triangles and crosses).  Each panel shows a different combination of observables;  upper left the average PSF ellipticity in both components; upper right, the average variation of the PSF ellipticity across the field of view; lower left a comparison of the average variation across the field of view to the PSF variation between the dithered image exposures of the field; lower right a comparison of seeing and airmass.  See text for the other combinations of parameters linked to the observations that were also found to show no clear trend between the accepted and rejected fields.}
   \label{fig:scatter}
\end{figure}

What can we learn from this set of fields where we are unable to adequately correct for the PSF distortion?  Figure~\ref{fig:scatter} shows just four examples of the many plots we inspected to compare how different observables related to the type of fields which passed or failed our systematics analysis.  Each point represents a different field with the tick mark indicating whether the field passed or failed the systematics analysis; open symbols indicate a field that has passed and crosses or filled circles indicate a field that has failed.  Before describing the figure in any detail, the key point to immediately note is that there is no combination of observables which separates the fields that fail from the fields that pass; the locus of the fields that pass is closely followed by the locus of the fields that fail.   This statement is true for all the different combinations of an extensive set of observables that were tested, as listed at the end of this section.  Focusing now on Figure~\ref{fig:scatter} in detail, the upper left shows the average PSF ellipticity in both components; upper right, the average variation of the PSF ellipticity across the field of view; lower left a comparison of the average variation across the field of view to the PSF variation between the dithered image exposures of the field; lower right a comparison of seeing and airmass. 

At the start of the CFHTLS observations the MegaCam PSF did not meet the design specifications but an unexpected solution was found by CFHT such that the image PSF became better behaved.  This solution involved inverting or `flipping' the third lens of the wide field corrector that is installed in front of the camera.  We can see the impact of this measure in the top right hand panel of Figure~\ref{fig:scatter} where we distinguish fields that were imaged before the CFHT `lens flip' (open and filled circles) and the majority of the data observed after the CFHT `lens flip' (open triangles and crosses).   This panel shows the variation of the PSF ellipticity across the field of view for both components for each field.  In this plane we can easily separate the data taken before the CFHT `lens flip' owing to its strong variation across the field of view in both ellipticity components.  What is interesting however is that we cannot separate the fields that fail and pass: the same fraction of fields fail before and after the CFHT `lens flip'.  This indicates that the spatial variation of our {\em lens}fit PSF model is sufficiently flexible to fit even the strongest PSF variation across the field of view.   

Our ability to correct PSF distortion is expected to depend on the amplitude of the ellipticity and size of the PSF.  The upper left panel of  Figure~\ref{fig:scatter} shows the distribution of fields as a function of the average PSF ellipticity in both components.  The majority of the data has PSF distortions within a few percent, with a slightly stronger $e_1^\star$ component.  The lower right panel of Figure~\ref{fig:scatter} shows the distribution of fields as a function of seeing (PSF size) and airmass.  All the $i'$ data have sub-arcsecond seeing with the CFHTLS seeing criterion of better than 0.8 arcsec met for the majority of the data.  There are a handful of poorer seeing fields, however, where weather conditions deteriorated during the exposure, for example,  in addition to early data taken before the CFHT `lens flip' whilst the survey queue scheduling system was optimised.  The survey was observed at low airmass as CFHT does not have an atmospheric distortion corrector.  As before, we see very little to distinguish fields that pass and fail in this space.  It is clear that {\em lens}fit is able to produce robust results for even the strongest PSF distortions with $|e^\star|\sim 0.1$ and the poorest of seeing, yet even with good seeing conditions and percent level PSF distortion our analysis can fail.  For the small sample of fields with seeing better than $0.6$ arcsec all fields pass.  This is a small sample so we are unable to make a conclusive statement on this, but it is gratifying to see that nothing in the analysis appears to compromise our ability to analyse the best quality data.

We now turn to the lower left panel of Figure~\ref{fig:scatter} which compares the average variation of the PSF ellipticity across the field of view to the average variation of the PSF ellipticity between the exposures.  Again we find nothing to indicate why fields pass and fail, but this plot illustrates the importance of analysing data from individual exposures compared to a stack, as it shows the variance of the PSF between the exposures is as significant as the variation of the PSF across the field of view.  This means that the PSF in a co-added image stack is a complex blend of varying elliptical profiles, breaking the typical assumptions about PSF profiles used in shape measurement methods.
{\em Lens}fit is the first method to optimally take into account the PSF variation between the exposures in a simultaneous analysis of the unstacked data, making it a significant advance in shear measurement.

Figure~\ref{fig:scatter} shows just four example plots to demonstrate the properties of the CFHTLenS data set.  We have also investigated potential correlations of the fields that pass and fail with wind speed, wind direction relative to the telescope dome slit, dome and CCD temperature, the number of stars, the area of the field lost to cosmic rays and defects, the fraction of area masked due to satellites or bright stars, the variance of the PSF size across the field of view and between exposures, the number of dithered exposures, higher order moments of the PSF and the Strehl ratio of the PSF, as estimated from the fraction of light in the central pixel.  We find no clear indication of a linear combination of observing conditions that causes some fields to fail our analysis.  This conclusion led us to look at random effects such as higher order correlations in the PSF distortion caused by atmospheric turbulence that we would be unable to model owing to the limited stellar number density.  This study is detailed in \citet{HeyRowe} which concludes that for the 600 second CFHTLenS exposure times, the high order turbulent correlations in the PSF are negligible.

\section{Systematic error analysis of two-point cosmic shear statistics}
\label{sec:2pt}
\begin{figure}
   \centering
   \includegraphics[width=3.5in, angle=0]{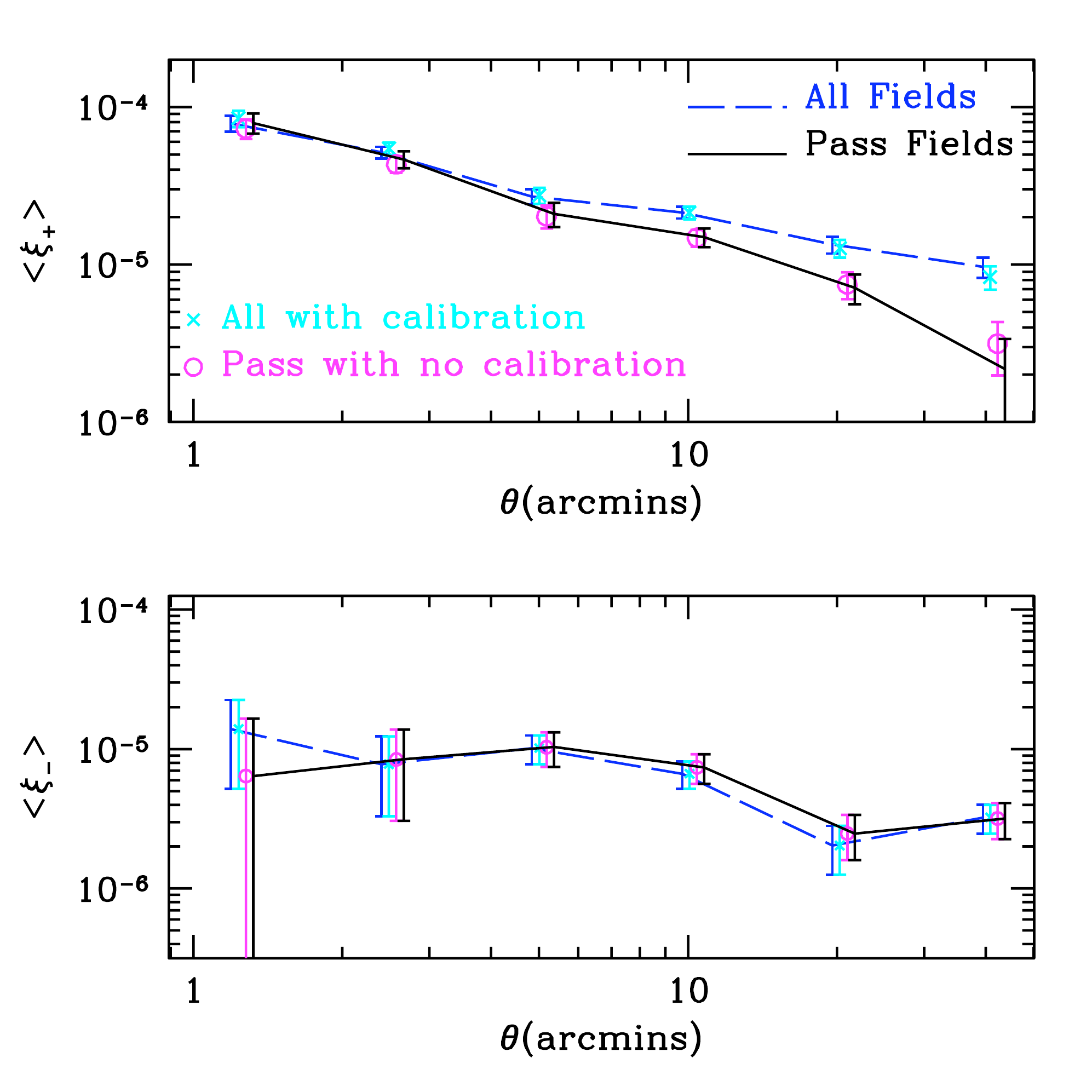} 
   \caption{Comparison of the two-point shear correlation statistic $\xi_+$ (upper) and $\xi_-$(lower) measured using all the data (dashed) and only the 75\% of the fields which pass our tests.  The impact of the application of the calibration correction detailed in \citet{Miller2012} is shown to be within the statistical errors.}
   \label{fig:xicomp}
\end{figure}

In the previous section we identified a small fraction of data which failed our criteria for cosmic shear analysis and we now assess the impact of this field selection on
the two-point shear correlation statistic $\xi_{\pm}$ (equations~\ref{eqn:xipm_est} and~\ref{eqn:xipm}) which we use to place tight constraints on cosmological parameters in \citet{Kilbinger2012}.    Figure~\ref{fig:xicomp} compares the two correlation functions as measured from the full survey area (dashed) and as measured from the survey area which passes our systematics tests (solid).  We can see that the field selection has a significant impact on the amplitude of the signal at large scales in $\xi_{+}$ and that the calibration corrections applied to the data, described in Section~\ref{sec:empcor} (crosses in the full survey area case and circles in case of the fields that pass) make no significant difference to the measured signal.  Note that the errors come from a bootstrap analysis.

\begin{figure}
   \centering
   \includegraphics[width=3.5in, angle=0]{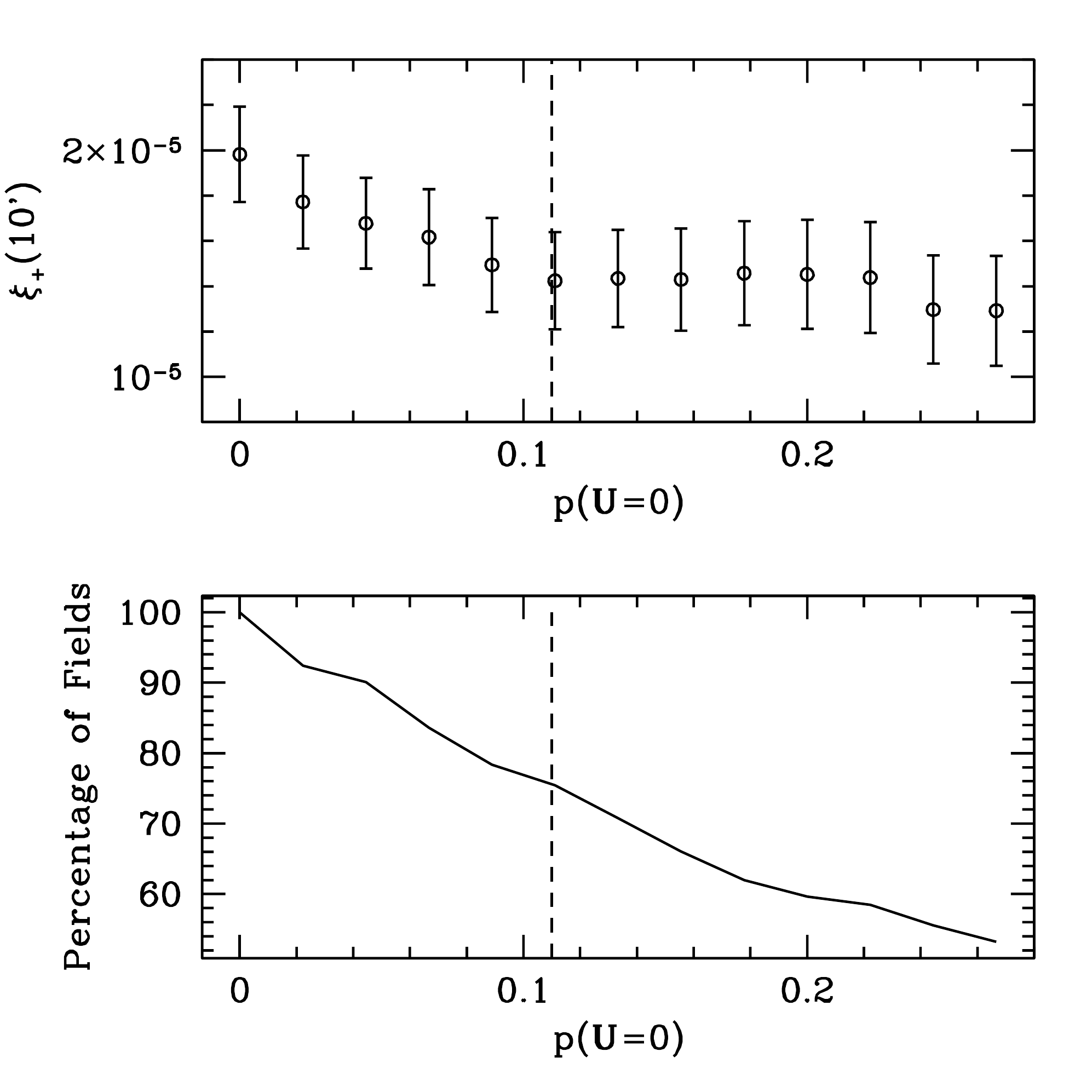} 
   \caption{Upper Panel: comparison of the two-point shear correlation statistic $\xi_+$ measured at 10 arcmin for varying systematics thresholds on $p(\bm{U}=0)$.  The percentage of fields that remain at each threshold is shown in the lower panel. Note that the errors are correlated as they show the field-to-field variance between the fields used at each data point.  In addition they do not include the increasing sampling variance error as the number of fields decrease.   The dashed lines indicate the optimal threshold that is defined independently, in Section~\ref{sec:fieldsel}.}
   \label{fig:xivarpcut}
\end{figure}

It is crucially important at this point of the analysis not to introduce confirmation bias into our cosmology analysis. This would occur for instance if we would assess the performance of our systematics by comparing our results to those we expect from parameter constraints from other surveys.  
The field selection and the threshold optimization to set $p(\bm{U}=0)>0.11$, described in section~\ref{sec:fieldsel}, was therefore made independently and blindly to any knowledge of its impact on the cosmological parameter constraints being investigated independently within the team.  To confirm the robustness of the systematics selection we developed an independent test, making progressively more conservative cuts to the data by increasing the threshold set on $p(\bm{U}=0)$, the probability that the fields had $\Delta \xi_{\rm obs}$ consistent with zero (see Section~\ref{sec:res}).  This 
is demonstrated in Figure~\ref{fig:xivarpcut} which shows, in the upper panel, the two-point shear correlation statistic $\xi_+$ measured at 10 arcmin using all data with a measured $p(\bm{U}=0)$ above the varying threshold shown on the horizontal axis.  The percentage of fields that remain at each threshold is shown in the lower panel.  This figure shows a progressive decrease in the amplitude of the large scale $\xi_+$ signal as fields with systematic errors are progressively removed from the analysis with the increasing cut on $p(\bm{U}=0)$.  This decrease in amplitude continues until the optimal threshold of $p(\bm{U}=0)>0.11$, set using a completely different and independent method in Section~\ref{sec:fieldsel}, is reached (shown dashed).  Removing more fields beyond that threshold gradually increases the noise on the measure, as the survey area decreases,  but does not decrease the amplitude of the large scale power any further.  This test gives us confidence that our selection criteria, which we define independently in comparison to simulations, is optimal and unbiased within the statistical noise of the survey.

\subsection{Field-to-field variation of large-scale cosmic shear signal}
\begin{figure}
   \centering
   \includegraphics[width=3.5in, angle=0]{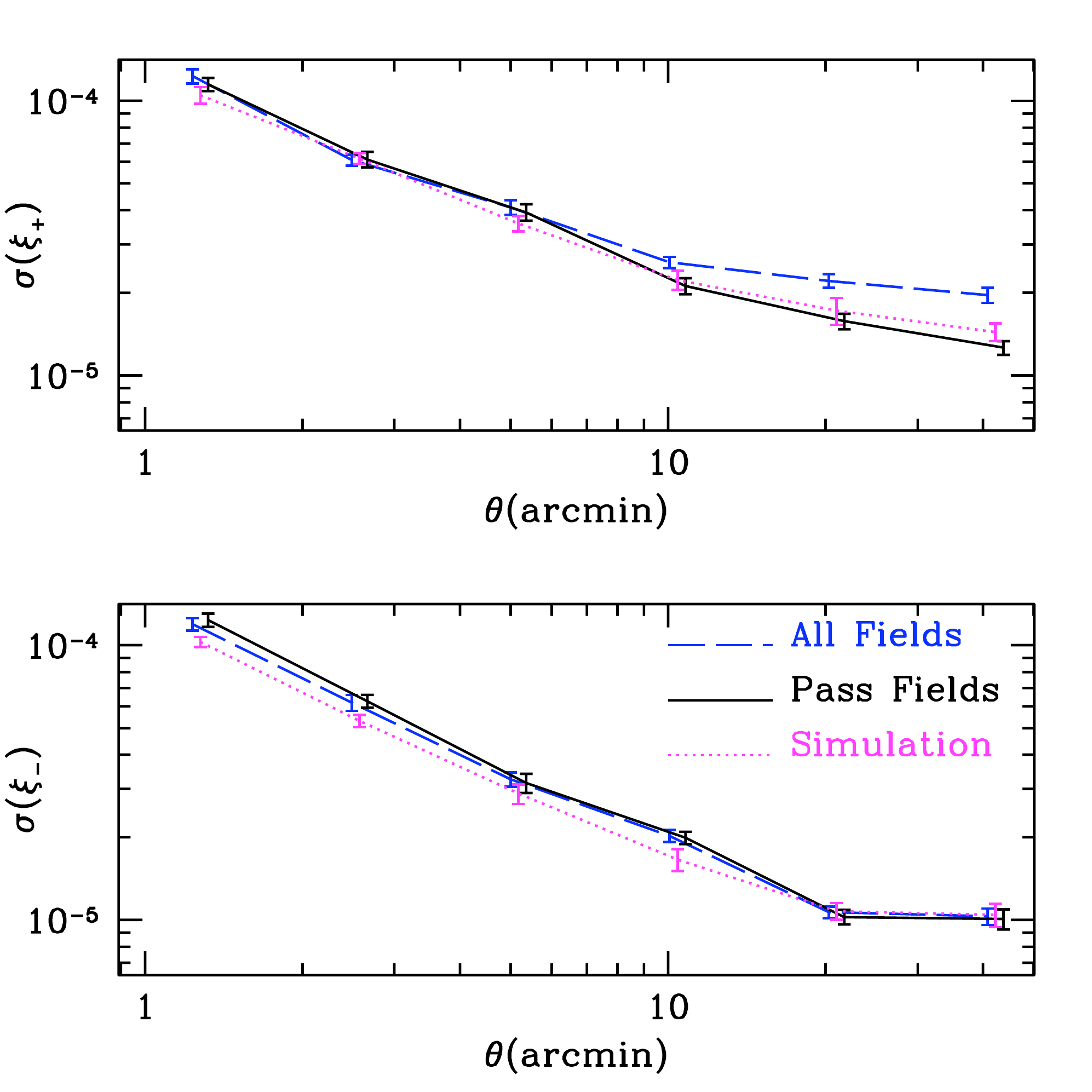} 
   \caption{Comparison of the variance between fields of the two-point shear correlation statistic $\xi_+$ (upper) and $\xi_-$(lower).  Using all the data reveals a significant increase in the variance between fields on large scales.  Using only the 75\% of the fields which pass our tests we find a field-to-field variance that is consistent with the sampling variance expected from a {\it  WMAP5} cosmology.}
   \label{fig:large_scale_variance}
\end{figure}

One of the first indications of systematic errors in the catalogues used in \citet{Fu08} came from an analysis of the field-to-field variation on large scales \citep{MK09}.    This is a cosmology-dependent test which we can not, and do not, use for field selection.  As we do not use the covariance of the two-point correlation function to estimate cosmological parameters in any future analyses however, we are afforded the opportunity, at this final stage, to compare the field-to-field variance measured to that expected for a {\it WMAP5} cosmology.  In Figure~\ref{fig:large_scale_variance} we therefore repeat the \citet{MK09} field-to-field variation analysis on the CFHTLenS data, showing the variance of the two-point shear statistics $\xi_{\pm}$ measured between the fields as a function of angular scales where the errors shown come from a bootstrap analysis.    The variance between the fields for the full survey (dashed) can be compared to the variance between the fields that pass our systematics tests, showing that the field selection brings down the variation between the fields significantly.  In addition the field-to-field variance can be compared to the variance expected from a {\it WMAP5} cosmology as predicted by an analysis of the CFHTLenS `clone' (see Section~\ref{sec:clone}), which is shown (dotted) to agree very well with the measurements.

\subsection{Analysis of tools previously advocated to detect systematic errors}
\label{sec:csysB}
\begin{figure}
   \centering
   \includegraphics[width=3.5in, angle=0]{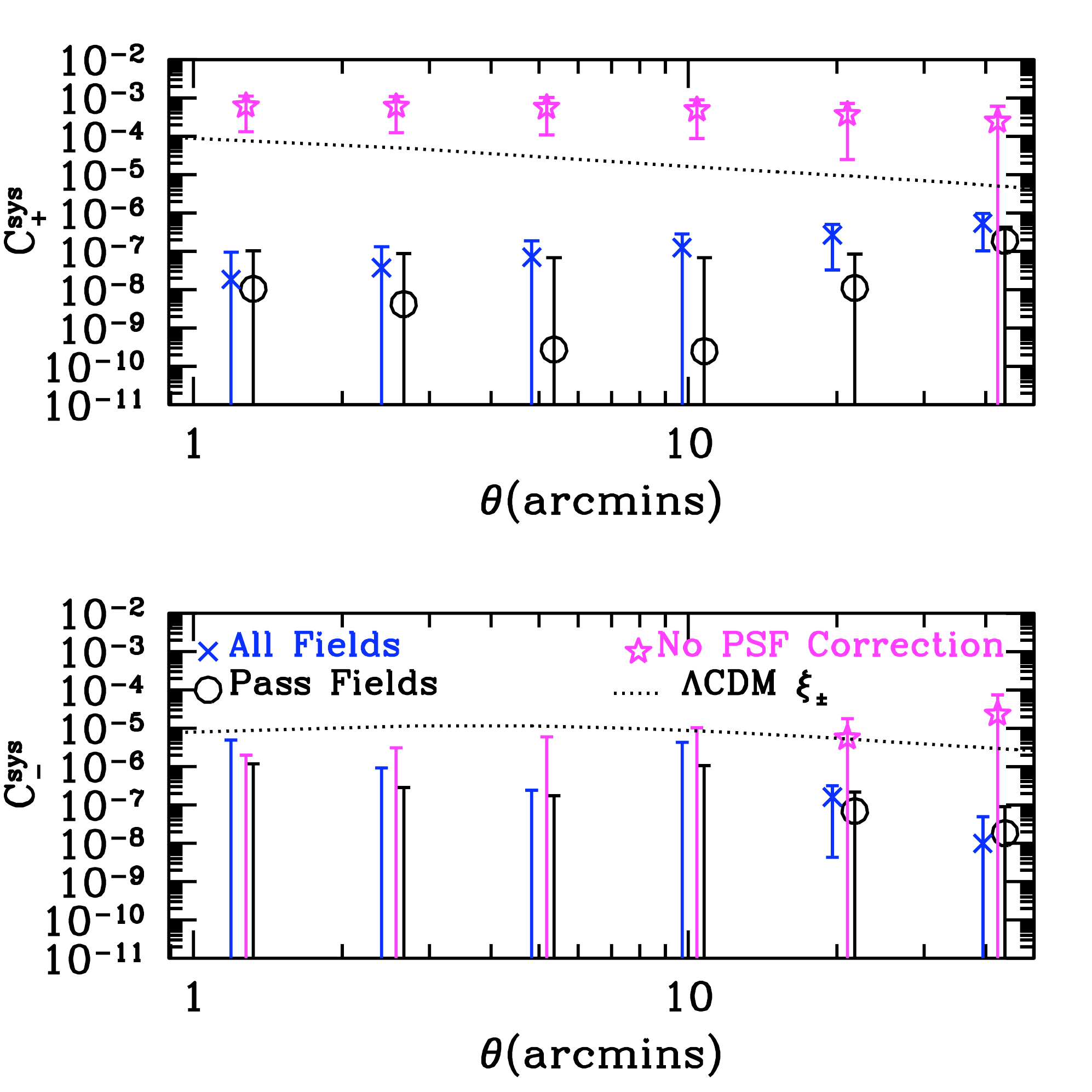} 
   \caption{Comparison of the $C^{\rm sys}_{\pm}$ statistic often used to test systematics, as measured on the full data set (crosses) and the fields which passed our systematics tests (circles).     The measured systematic signal can be compared to the lensing signal we would expect to measure for a {\it  WMAP5} cosmology (shown dashed).   We find that even when using the full sample we find a signal for $C^{\rm sys}_{\pm}$ that is consistent with zero, concluding that the $C^{\rm sys}_{\pm}$ statistic does not provide a very sensitive test of the levels of systematics in the CFHTLenS survey.  The average CFHT MegaCam PSF correlation function $\langle e^\star e^\star \rangle$ is also shown for comparison (stars)}
   \label{fig:csys}
\end{figure}

In previous weak lensing analyses, two standard tools have been used to infer a level of systematics in the data that is consistent with zero.  The first test is essentially $\Delta \xi_{\rm obs}(\theta)$ (equation~\ref{eqn:deltaeobs}) but averaged over the full survey, and using an average PSF from the stack.  This is often written as
\be
C^{\rm sys}_{\pm} (\theta) = \frac{\langle \epsilon^{\rm obs} e^\star \rangle^2}{\langle e^\star e^\star \rangle} \, ,
\ee
\citep{Baconcsys} where the correlations are calculated following equation~\ref{eqn:xipm_est}.  Figure~\ref{fig:csys} compares this statistic as measured from the full CFHTLenS survey (crosses) and the selected areas of the survey that pass our systematics tests (circles).   
The results show both measures to be consistent with zero and well below the amplitude of the lensing signal we would expect to measure for a {\it  WMAP5} cosmology (shown dashed).  
Note that we choose to use a log-scale to demonstrate how small this measured systematic signal is, at the expense of losing negative data points which are all consistent with zero signal.
With this test alone we might incorrectly conclude our full survey was systematics free.   In this full survey case, we suspect that the errors are averaging out when this statistic is measured across the survey.  This therefore makes it crucially important to undertake systematics error testing on a field by field and exposure by exposure basis, as we have with our measurements of $\Delta \xi_{\rm obs}$ (equation~\ref{eqn:deltaeobs}).

Figure~\ref{fig:csys} also shows the CFHT MegaCam PSF correlation function $\langle e^\star e^\star \rangle$ (stars) which has, on average, zero signal in the $\xi_{-}$ correlation (lower panel) compared to a significant essentially constant signal in the $\xi_{+}$ correlation (upper panel).  In Figures~\ref{fig:xicomp} and~\ref{fig:large_scale_variance} we see that the impact of our field selection is always in the $\xi_+$ statistic. This therefore suggests that at least part of the systematic signal in our failed fields is related to the PSF.  This could potentially arise from PSF modeling errors or as a result of noise bias in the likelihood surfaces as discussed previously in Section~\ref{sec:Anoise}. 

\begin{figure}
   \centering
   \includegraphics[width=3.3in, angle=0]{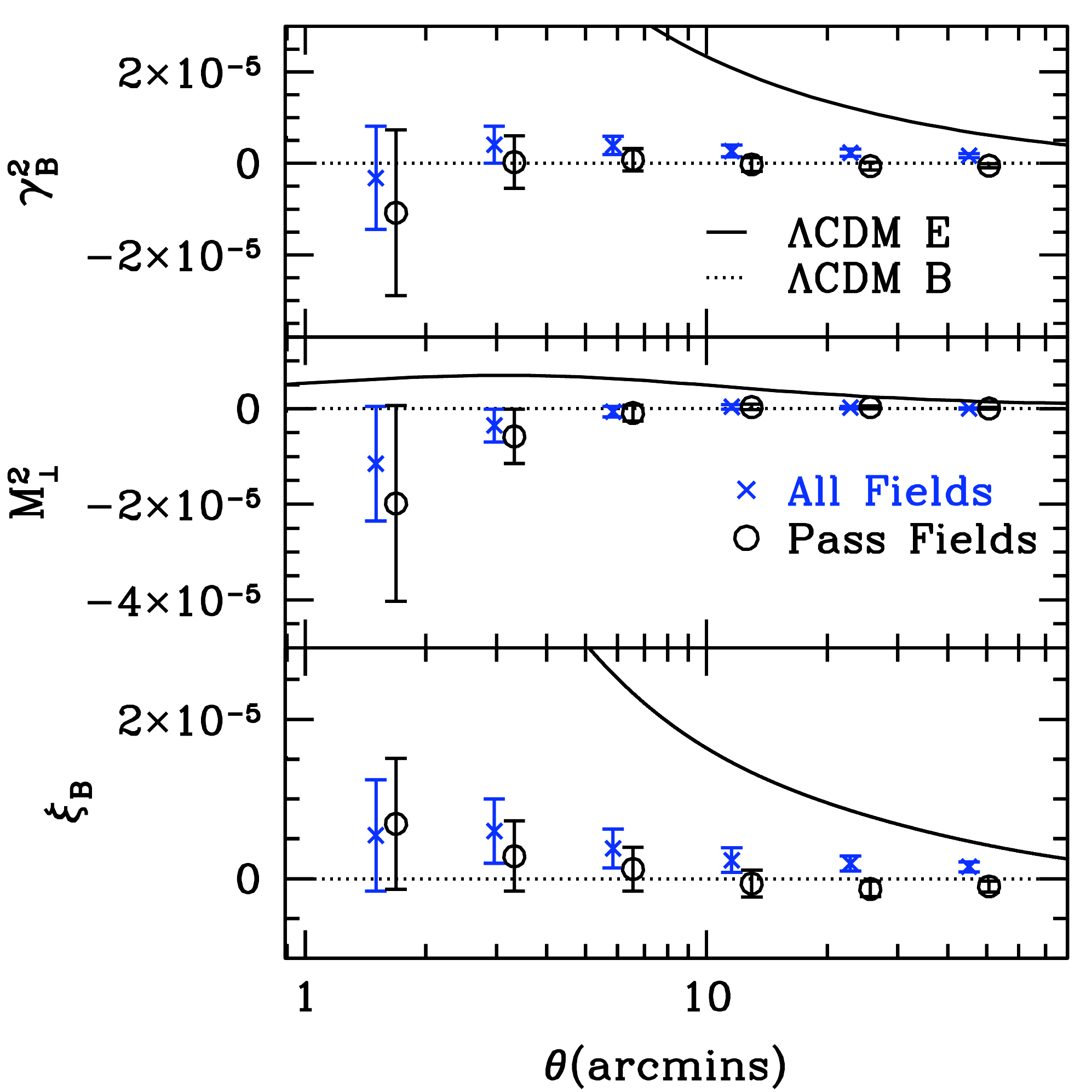} 
   \caption{The B-mode signal measured using three different two-point statistics;  the top-hat shear statistic $\gamma^2_B$ (upper panel), the mass aperture statistic $M_\perp^2$ (middle panel), and the two-point shear correlation function $\xi_B$ (lower panel).  The B-modes measured from the fields that pass our systematics tests (circles) can be compared with those measured from all fields (crosses) and the E-mode amplitude of the lensing signal that we would expect to measure for a {\it  WMAP5} cosmology (shown solid).}
   \label{fig:Bmode}
\end{figure}

The second standard weak lensing systematics test is an E/B mode decomposition of the signal.  Cosmic shear produces an almost pure E-mode only distortion pattern, with some small B-mode signal expected at angular scales less than a few arcminutes, resulting from source redshift clustering \citep{SchvWKM02}.   A detection of a significant B-mode indicates the presence of a non-lensing systematic signal in the data.  As the CFHTLenS PSF has no $\xi_{-}$ correlation on average, the B-mode caused by any PSF residuals will equal the E-mode caused by the same PSF residuals.  In this particular case, therefore, an E/B mode decomposition is effective at detecting a PSF-related systematic.  One drawback however is that to avoid bias in the E/B measurement from data,  the most typically used E/B mode decomposition method from \citet{SchvWKM02} formally requires an input cosmology which we generally wish to avoid using in our blind systematics tests \citep[see for example][]{KSE06}.   Alternative E/B decomposition statistics have been proposed to avoid this cosmology dependence, \citep{SK07,COSEBIS}, and we will investigate these statistics in future work.    Figure~\ref{fig:Bmode}  compares the B-mode signal measured from the data using three different two-point statistics;  the top-hat shear statistic $\gamma^2_B$ (upper panel), the mass aperture statistic $M_\perp^2$ (middle panel), and the two-point shear correlation function $\xi_B$ (lower panel).  These were calculated by integrating over the two-point shear correlation function $\xi_{\pm}$, measured from the data in 2000 fine angular bins between $0.13<\theta<70$ arcmin, using the decomposition filter functions in equations 26, 32, 41 of \citet{SchvWKM02}.  We calculate the required cosmology dependent normalization by extending the measured $\xi_{\pm}$ using a {\it  WMAP5} cosmology model over the small and large angular scales that are inaccessible to the CFHTLenS data  \citep[see][for more detail on the implementation of this correction]{HymzGEMS}.  The B-modes measured from the fields that pass our systematics tests (circles) are consistent with zero on all scales.  This can be compared with the B-modes measured from all fields (crosses) which are found to be significant on some scales for some of the statistics tested.   In the case of the B-mode of the two-point shear correlation function $\xi_B$ (lower panel) we clearly see an excess large-scale B-mode signal when all the data is analysed.  This signal is of similar amplitude to the change in the measured two-point shear correlation function $\xi_+$, shown in Figure~\ref{fig:xicomp}, when the failed fields are removed from the analysis.  We note however, that the B-modes for all fields, whilst significant, are well below the amplitude of the lensing signal that we would expect to measure for a {\it  WMAP5} cosmology (shown solid).  
Had only this B-mode test been performed, one might therefore conclude that shear systematic errors were tolerable.  Yet in Figure~\ref{fig:xicomp} the difference made by the field selection of Section~\ref{sec:fieldsel} can be clearly seen.

\subsection{The impact of using automated or manual masks for cosmic shear}
\label{sec:masks}
The importance of correctly masking data to remove image artifacts has been long understood in all applications of astronomy, but it is of particular importance for weak lensing analyses.  Even a small fraction of diffraction spikes or broken detections along long satellite trails erroneously present in a galaxy catalogue can introduce a strong systematic signal as they align.  The {\sc THELI} reduction pipeline includes an automated masking routine to detect for such artifacts \citep[see][for more details]{Erben12}.  Using the $i'$-band stacked data, we manually inspect and modify these candidate masks  for use in our final analysis.  In the majority of the cases any modifications made are because the automated routine is being too conservative and the manual inspection recovers back $\sim1$\% of the valuable area masked unnecessarily.  Manual inspection also detects the occasional faint satellite trail and missed faint diffraction spike by {\sc THELI} and has the added benefit of ensuring that at least one person has inspected each image for any unusual behaviour.  The value of this effort however has to be balanced by the number of human hours invested in such an endeavor, which for CFHTLenS consisted of roughly 20 minutes per field and hence a total of $\sim 60$ hours for the full survey.   In addition, whilst every effort was made to maintain a quality control by comparing two independent sets of masks for a sub-set of fields and running the majority of fields through a second-pass quality assurance check, the nature of visual inspection is, by definition, subjective and therefore inherently dependent on the team members who performed the task.

\begin{figure}
   \centering
   \includegraphics[width=3.5in, angle=0]{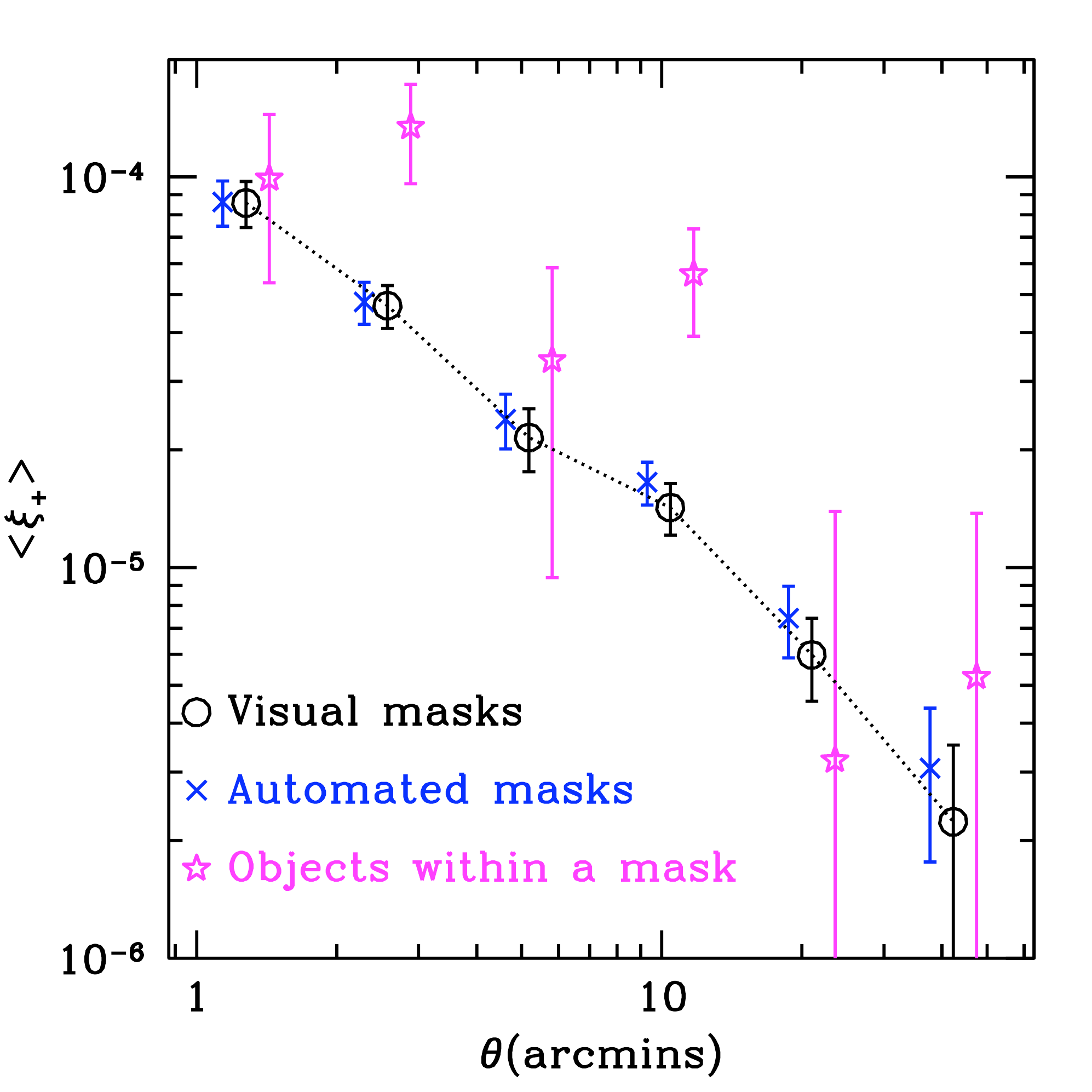} 
   \caption{The two-point shear correlation statistic $\xi_+$ measured using the sample of data that passes our systematics tests masked in three different ways; galaxies outside the final manually improved masks (circles), galaxies outside the automated initial candidate masks (crosses) and galaxies that lie within a masked region (stars).  The automated and mask data points (crosses and stars) are offset in each $\theta$ direction for clarity.}
    \label{fig:maskcomp}
\end{figure}

We now quantify how important this manual masking step is in Figure~\ref{fig:maskcomp}, by comparing the two-point shear correlation statistic $\xi_+$ measured using the sample of data that passes our systematics tests, masked in three different ways; galaxies outside the final manually improved masks (circles, as used in all analyses so far), galaxies outside the initial automated candidate masks (crosses) and galaxies that lie within a masked region (stars).  The results show that for the statistical accuracy achieved by CFHTLenS, the signal using the candidate mask and manual mask are consistent.    We should note that in contrast to previous lensing surveys, CFHTLenS applies a model fitting method to the data.  Any objects that appear to have a complex morphology, or those that are very poorly fit by a two-component galaxy model are therefore given a zero weight and thus rejected from the sample.  As such {\em lens}fit is providing an alternative for the removal of any artifacts missed by the automated masking routine.    The strong signal from the objects within the masked region does however show the importance of removing regions of image artifacts and that using a model fitting analysis alone is not sufficient.  This result is very promising for the use of automated masking in future surveys.  When survey statistical power grows significantly compared to CFHTLenS however, the robustness of automated masking should be re-assessed through an extensive series of image simulation analyses.

\section{Testing Redshift Scaling with Galaxy-galaxy lensing}
\label{sec:zscale}
In \citet{MK09}, a preliminary tomographic analysis of the CFHTLS found a cosmic shear signal for galaxy redshifts $z > 1$ that was inconsistent with the signal expected from the best fitting $\Lambda$CDM cosmology as measured from a purely 2D analysis of the data in \citet{Fu08}.  This was seen as an indication of systematic biases suggesting either an error with the photometric redshifts used at that time or a redshift dependent shape measurement bias.  In light of this earlier finding we require a final test of the robustness of the CFHTLenS data to the variation in lensing signal with redshift.
In this section we therefore develop and implement a redshift scaling test for the data.  We are unable to use the cosmology dependent redshift testing methodology of \citet{MK09}, as we use the CFHTLenS tomographic signal to constrain cosmological parameters in \citet{Benjamin2012}.  Our method therefore instead considers the redshift evolution of the tangential ellipticity $\epsilon_{\rm t}(\theta)$ measured about a lens at angular separation $\theta$ from a source with complex ellipticity $\epsilon$ (equation~\ref{eqn:e1e2}).  This is commonly known as galaxy-galaxy lensing whose redshift evolution is only very weakly sensitive to our assumptions on the fiducial cosmology, in contrast to the evolution of the cosmic shear signal which is very sensitive to a change in cosmology.

The mean tangential shear around a sample of lens galaxies can be approximately described by a singular isothermal sphere (SIS) shear profile,
\citep{2001A&A...378..361B} 
\begin{equation}
  \overline{\gamma}_{\rm t}^{(ij)}(\theta)=
  \frac{2\pi}{\theta}\left(\frac{\sigma_{\rm v}}{c}\right)^2
  \Ave{\frac{D_{\rm ls}}{D_{\rm s}}}_{ij}\;,
\end{equation}
where $\sigma_{\rm v}$ is the characteristic velocity of the SIS, $c$ is the speed of light, and the ratio of angular diameter distances 
\begin{eqnarray}
  \label{eq:shearratio}
  \lefteqn{\Ave{\frac{D_{\rm ls}}{D_{\rm s}}}_{ij}=}\\
  &&\nonumber
  \int_0^{z_{\rm max}}\!\!\!\d z_{\rm l}\,
  p^{(i)}_{\rm l}(z_{\rm l})
  \int_{z_{\rm l}}^{z_{\rm max}}\!\!\!\d z_{\rm s}\,
  p^{(j)}_{\rm s}(z_{\rm s})\,
  \frac{f_{\rm K}\Big(w(z_{\rm s})-w(z_{\rm l})\Big)}
  {f_{\rm K}\Big(w(z_{\rm s})\Big)} \, ,
\end{eqnarray}
is a purely geometrical factor that depends on the effective redshift probability
distribution function $p^{(i)}_{\rm l}(z)$ of lenses and
$p^{(j)}_{\rm s}(z)$ of sources. $f_{\rm K}(w)$ denotes the co-moving angular
diameter distance for a co-moving radial distance $w$ and curvature $K$
of the fiducial cosmological model
\citep[see for example][]{Bible}. $z_{\rm max}$ is the maximum observed galaxy redshift beyond which
$p_{\rm s}(z)$ is zero.  
For a given lens subsample in a fixed redshift bin $i$, the mean tangential shear of galaxies in redshift bin $j$  increases with mean source redshift of the bin. 
The relative change of the shear signal (for a fixed lens redshift) is a very weak function of the cosmological
parameters, such that equation~\ref{eq:shearratio} robustly predicts the
relative change of $\overline{\gamma}_{\rm t}^{(ij)}(\theta)$ for a wide range of cosmological parameters. The absolute amplitude of $\overline{\gamma}_{\rm t}^{(ij)}$ for a fixed $i$ is given by the SIS velocity $\sigma_{\rm  v}$ which will vary for each lens sample. 
We choose to use the simple SIS profile for the purposes of this test as this has been shown to be a good model for the data for angular scales $\theta <3$ arcmin \citep{RCS2}. 

We select a sample of lens galaxies with magnitude $i'_{\rm AB} <22.5$ 
and the full CFHTLenS catalogue for our source sample.
All galaxies have a photometric redshift estimate and associated error distribution $P(z)$ \citep{HH12}.  We use these to initially split both samples into $N_{\rm z}=6$
photometric redshift bins; $[0.0,0.2]$, $[0.2,0.4]$, $[0.4,0.5]$, $[0.5,0.7]$, $[0.7,0.9]$, and
$[0.9,1.3]$, and then calculate the effective redshift distribution for each bin from a weighted average of the probability distributions of the galaxies within each bin, incorporating each source galaxy weight.
For each of the $N_{\rm z}^2$ combinations of lens $i$ and source $j$
subsamples we compute the mean tangential shear
$\overline{\gamma}^{(ij)}_{\rm t}(\theta)$
by averaging source ellipticities $\epsilon_{\rm t}$ around the lenses in bins with an angular scale ranging
from 0.8 arcmin to 4 arcmin.   As we are interested in testing redshift scaling we measure $\epsilon_{\rm t}$ for sources that are both in front of and behind the lenses.  We take a weighted average using the stellar mass of the lens,
estimated photometrically from the data \citep{Velander2012}.  This weighting scheme aims to optimize the signal-to-noise of the measurement, as the stellar mass is expected to correlate with halo mass. The measurement is made for each field which passes our systematics tests in Section~\ref{sec:fieldsel} and we make a Jackknife statistical error estimate of the mean signal over the survey \citep{1986ANNSTAT}. 

\begin{figure}
   \includegraphics[width=3.35in, angle=0]{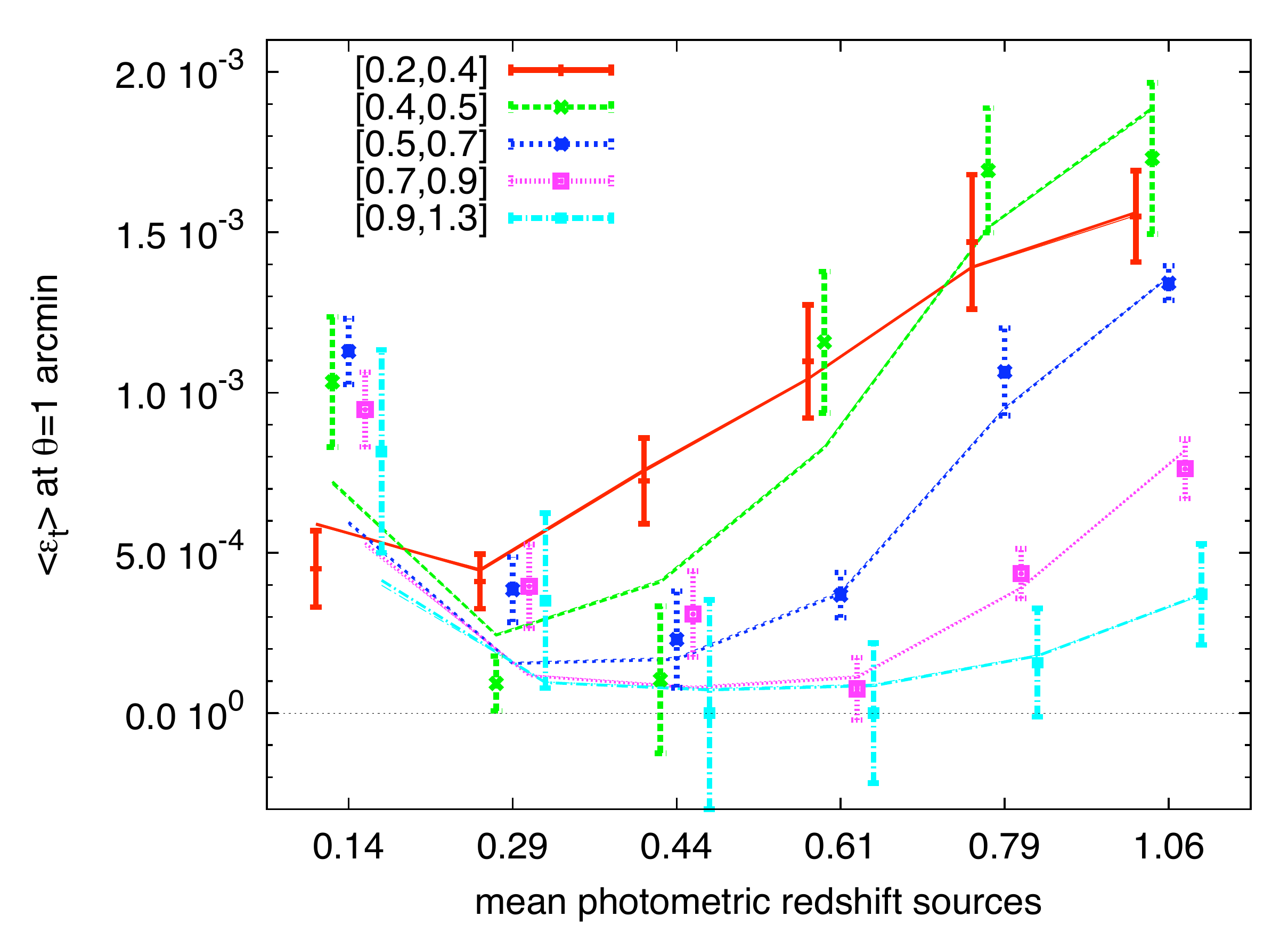} 
   \caption{The mean tangential shear at an angular
    separation of one arcmin as function of the weighted mean source photometric redshift. 
    The data points are obtained from the
    best-fitting SIS model of the data between 0.8 and 4 arcmin. Different symbols correspond
    to different redshift lens subsamples (see legend). The lines correspond to a {\it WMAP5} cosmology
    theoretical prediction of the change in the shear signal with an
    overall amplitude that is fixed by the best-fitting SIS
    velocity and allowing for the full $P(z)$ distribution of the source redshifts. This yields a non-zero expectation for sources whose nominal photometric redshift places them in front of the photometric redshift of the lenses.}
   \label{fig:zscaling}
\end{figure}

Figure~\ref{fig:zscaling} shows the amplitude of the best-fitting SIS model for
$\overline{\gamma}_{\rm t}^{(ij)}$ at $\theta=1$ arcmin 
as a function of the mean photometric redshift of the sources for the six photometric redshift bins.   The data corresponding to each of the lens redshift bins is
indicated in the legend, where for clarity we only plot the data from the source-lens bin combinations where the lens redshifts are 
in the trusted $0.2<z<1.3$ redshift range (see section~\ref{subsec:photom}).
The lines are trends
expected from a {\it WMAP5} cosmological model, incorporating the
redshift probability distribution $P(z)$ of lenses and sources as determined from the data.   The amplitudes of the expected curves are adjusted
by finding the value of $\sigma_{\rm v}$ that best fits the data for
each lens subsample using all sources. For higher mean lens
redshifts, the SIS velocity increases as expected for a flux-limited
sample and we refer the reader to \citet{Velander2012} for a detailed galaxy-galaxy lensing 
analysis to measure halo mass in CFHTLenS.

The most striking feature of Figure~\ref{fig:zscaling} is the significant signal detected from source galaxies with photometric redshifts $z<0.2$.  If the photometric redshift measure was accurate we would not expect to measure a galaxy-galaxy lensing signal for any of the lens samples shown, as these sources would be in front of all the lenses.  It is most likely that this signal results from the strong catastrophic errors measured for $z<0.1$ in \citet{HH12}, where high redshift sources have been misclassified at low redshifts.  The rise in the expected {\it WMAP5} cosmological model for this low redshift source bin shows that some of this catastrophic error information is contained in the photometric redshift probability distribution $P(z)$ which is used to compute the model.  But the data for this source redshift bin remains inconsistent with the models for the majority of lens samples, supporting our choice not to trust the data with photometric redshift estimates $z<0.2$.

With the exception of the data with $z<0.2$, we find that we measure a redshift dependence of the  galaxy-galaxy lensing signal that is consistent with a cosmological model for redshifts $0.2<z<1.3$.  Figure~\ref{fig:zscaling} therefore demonstrates the robustness of the CFHTLenS shear and redshift catalogues by confirming the absence of any significant redshift dependent biases in our method.  The excellent agreement at the high redshifts that were contaminated by systematic errors in an earlier CFHTLS analysis, \citep[see][for details]{MK09}, is particularly encouraging.
Note that assuming an open cold dark matter model with $\Omega_\Lambda = 0$ produces a theoretical curve that is indistinguishable from the $\Lambda$CDM curve shown on Figure~\ref{fig:zscaling}, clarifying that this test is insensitive to the assumed fiducial cosmology.  We therefore conclude that once the calibration and selection procedures outlined in this paper are incorporated, our weak lensing catalogues are of a very high quality and ready for the application of many varied cosmological studies.

\section{Conclusions}
\label{sec:conc}
This paper presents the Canada-France-Hawaii Telescope Lensing Survey (CFHTLenS) and is one out of four technical papers detailing and verifying the accuracy of each component of the CFHTLenS data analysis pipeline which is overviewed in Section~\ref{sec:pipe}.  We dissect the pixel level analysis in \citet{Erben12}, the photometry and redshift analysis in \citet{HH12}, and the shear and image simulation analysis in \citet{Miller2012}.  In this paper we focus on bringing all these key stages of the pipeline together to perform the final step of the analysis, identifying and removing any problematic data which contaminates the survey.   As illustrated in Table~\ref{tab:comparison}, the complete pipeline has been re-written for CFHTLenS as every core stage of the standard pipeline used in previous analyses could have introduced low-level systematic errors.  We therefore hope that the set of CFHTLenS technical papers will provide a detailed record of this work, acting as a reference to aid the future analyses of upcoming weak lensing surveys.  We argue that for a science goal that is as demanding on the data analysis as the study of weak gravitational lensing, every pipeline stage must be integrated and optimised for shape and photometry measurements.   From the raw data to the shear and photometric redshift catalogues there are many stages where systematic errors can be introduced. As such we advocate that surveys take a holistic approach to their lensing analyses as we have for CFHTLenS.

One of the most important developments during the CFHTLenS project resulted from using the {\em lens}fit shape measurement method's ability to optimally analyse individual exposures in comparison to stacked data.  This improved the quality of the PSF model and hence the accuracy of the resulting shear measurement.    That said, this particular advance cannot be identified as the sole solution to the systematic errors that concerned earlier analyses of CFHTLS \citep[see a discussion of these errors in][]{MK09}.  Indeed we conclude that it was most likely the combination of many small low-level effects, discussed in Section~\ref{sec:pipe}, that we have sought to identify at source and improve in this analysis.

Whilst every effort has been made to develop a perfect end-to-end analysis tool, and significant improvements have been made,  we still find ourselves in the unsatisfactory position of requiring the removal of 25\% of the observed fields from our analysis (see Section~\ref{sec:fieldsel}) in addition to the application of small but significant calibration corrections \citep[see][and Section~\ref{sec:empcor}]{Miller2012}.  This paper is primarily focused on the development (Section~\ref{sec:method}) and application (Section~\ref{sec:res}) of a method to both calibrate and identify problematic data.  By incorporating full cosmic shear N-body simulations we are able to accurately determine the noise levels in a field by field, exposure level star-galaxy cross correlation analysis, in order to identify fields with a significant PSF residual.   We have demonstrated how adept this new method is to identifying individual fields where PSF residuals contaminate the signal.  We have also shown how the resulting measured two-point cosmic shear signal is robust to the threshold criteria we set on this systematic test in Section~\ref{sec:2pt}.  For comparison purposes we present the results of two standard systematic tests for the data, demonstrating how survey average tests are less sensitive to systematic error signals which become diluted when considered in a survey average, but are still present at a local level within the data.   

In Section~\ref{sec:scatter} we presented a comprehensive search which found no evidence for any correlations between the 25\% of fields which failed our systematics analysis and parameters which characterized the CFHT dome, telescope, camera and atmospheric conditions at the time of the observations.  We are therefore unable to conclude whether this level of rejected data should be expected for all lensing surveys, all ground-based surveys, all {\em lens}fit analyses or all MegaCam observations.   The CFHTLenS pipeline is currently in use to re-analyse the CFHT Red Sequence Cluster Survey \citep[RCS2, see for example][]{RCS2}, and analyse the CFHT $i'$-band imaging survey of the SDSS Stripe 82 region, the Next Generation Virgo Survey \citep[NGVS,][]{NGVS}, and the 
Kilo Degree Survey (KiDS) on the VLT Survey Telescope.  It will be interesting to compare the relative performance of the pipeline between these different surveys.

We demonstrate that our shear and photometric redshift catalogues are `science-ready', with a final cosmology insensitive galaxy-galaxy lensing test in Section~\ref{sec:zscale}, to confirm that the survey is not subject to the strong redshift-dependent biases found in earlier analyses.    These catalogues are being used for a variety of lensing analyses including the first set of CFHTLenS cosmological parameter constraints \citep{Kilbinger2012,Benjamin2012} with constraints on parameterised models of modifications to gravity \citep{Simpson} and constraints on the connection between galaxy properties and their parent dark matter haloes \citep{Velander2012}.  
With the community embarking on three major new weak lensing surveys\footnote{KiDS: kids.strw.leidenuniv.nl, DES:  www.darkenergysurvey.org and HSC: www.subarutelescope.org/Projects/HSC}, this is a very exciting time for the study of the `Dark Universe' with weak gravitational lensing.   We hope that the suite of detailed CFHTLenS technical papers will serve as a solid core upon which to build the hyper-precision weak lensing analyses of the future.

\section{Acknowledgements}
We would particularly like to thank the dedication of the CFHTLenS manual masking team that included some of the co-authors in addition to Jonathan Benjamin, Raphael Gavazzi, Bryan Gillis, Emma Grocutt, Karianne Holhjem, Martha Milkeraitis and Merijn Smit.  We would also like to thank the anonymous referee for their prompt helpful comments and thorough reading of the manuscript, in addition to
Alan Heavens, Gary Bernstein, Mike Jarvis, Sarah Bridle and Robert Lupton for their many constructive ideas during the development of this project.

This work is based on observations obtained with MegaPrime/MegaCam, a joint project of CFHT and CEA/DAPNIA, at the Canada-France-Hawaii Telescope (CFHT) which is operated by the National Research Council (NRC) of Canada, the Institut National des Sciences de l'Univers of the Centre National de la Recherche Scientifique (CNRS) of France, and the University of Hawaii. This research used the facilities of the Canadian Astronomy Data Centre operated by the National Research Council of Canada with the support of the Canadian Space Agency.  We thank the CFHT staff for successfully conducting the CFHTLS observations and in particular Jean-Charles Cuillandre and Eugene Magnier for the continuous improvement of the instrument calibration and the {\sc Elixir} detrended data that we used. We also thank TERAPIX for the quality assessment and validation of individual exposures during the CFHTLS data acquisition period, and Emmanuel Bertin for developing some of the software used in this study. CFHTLenS data processing was made possible thanks to significant computing support from the NSERC Research Tools and Instruments grant program, and to HPC specialist Ovidiu Toader.  The N-body simulations used in this analysis were performed on the TCS supercomputer at the SciNet HPC Consortium. SciNet is funded by: the Canada Foundation for Innovation under the auspices of Compute Canada; the Government of Ontario; Ontario Research Fund - Research Excellence; and the University of Toronto.  The early stages of the CFHTLenS project were made possible thanks to the support of the European CommissionÕs Marie Curie Research Training Network DUEL (MRTN-CT-2006-036133) which directly supported six members of the CFHTLenS team (LF, HHi, PS, BR, CB, MV) between 2007 and 2011 in addition to providing travel support and expenses for team meetings.

CH, HHo \& BR acknowledge support from the European Research Council under the EC FP7 grant numbers 240185 (CH), 279396 (HHo) \& 240672 (BR).   LVW acknowledges support from the Natural Sciences and Engineering Research Council of Canada (NSERC) and the Canadian Institute for Advanced Research (CIfAR, Cosmology and Gravity program).  TE is supported by the Deutsche Forschungsgemeinschaft through project ER 327/3-1 and, with PS, is supported by the Transregional Collaborative Research Centre TR 33 - `The Dark Universe'. HHi is supported by the Marie Curie IOF 252760 and by a CITA National Fellowship. HHo also acknowledges support from Marie Curie IRG grant 230924 and the Netherlands Organisation for ScientiÞc Research grant number 639.042.814.   TDK acknowledges support from a Royal Society University Research Fellowship.  YM acknowledges support from CNRS/INSU (Institut National des Sciences de l'Univers) and the Programme National Galaxies et Cosmologie (PNCG).  LF acknowledges support from NSFC grants 11103012 and 10878003, Innovation Program 12ZZ134 and Chen Guang project 10CG46 of SMEC, and STCSM grant 11290706600.  MJH acknowledges support from the Natural Sciences and Engineering Research Council of Canada (NSERC).  TS acknowledges support from NSF through grant AST-0444059-001, SAO through grant GO0-11147A, and NWO. MV acknowledges support from the Netherlands Organization for Scientific Research (NWO) and from the Beecroft Institute for Particle Astrophysics and Cosmology.  CB is supported by the Spanish Science Ministry AYA2009-13936 Consolider-Ingenio CSD2007-00060, project 2009SGR1398 from Generalitat de Catalunya and by the European CommissionÕs Marie Curie Initial Training Network CosmoComp (PITN-GA-2009-238356).  Part of BR's work was done at the Jet Propulsion Laboratory, California Institute of Technology, under contract with NASA.

{\small Author Contributions: All authors contributed to the development and writing of this paper.  CH and LVW co-led the CFHTLenS collaboration.  The authorship list reflects the lead authors of this paper (CH, LVW and LM) followed by two alphabetical groups.  The first alphabetical group includes key contributers to the science analysis and interpretation in this paper, the founding core team and those whose long-term significant effort produced the final CFHTLenS data product.  The second group covers members of the CFHTLenS team who made a significant contribution to the project and/or this paper.}

\bibliographystyle{mn2e}
\bibliography{ceh_2010}

\begin{thebibliography}{79}
\expandafter\ifx\csname natexlab\endcsname\relax\def\natexlab#1{#1}\fi

\bibitem[{{Abazajian} {et~al}\mbox{.}(2009){Abazajian}, {Adelman-McCarthy},
  {Ag{\"u}eros}, {Allam}, {Allende Prieto}, {An}, {Anderson}, {Anderson},
  {Annis}, {Bahcall}, \& et~al.}]{SDSS-DR7}
{Abazajian} K.~N. {et~al.}, 2009, ApJS, 182, 543

\bibitem[{{Amara} \& {R{\'e}fr{\'e}gier}(2008)}]{AmaraRef08}
{Amara} A., {R{\'e}fr{\'e}gier} A., 2008, MNRAS, 391, 228

\bibitem[{Bacon {et~al}\mbox{.}(2000)Bacon, Refregier, \& Ellis}]{BRE}
Bacon D., Refregier A., Ellis R., 2000, MNRAS, 318, 625

\bibitem[{{Bacon} {et~al}\mbox{.}(2003){Bacon}, {Massey}, {Refregier}, \&
  {Ellis}}]{Baconcsys}
{Bacon} D.~J., {Massey} R.~J., {Refregier} A.~R., {Ellis} R.~S., 2003, MNRAS,
  344, 673

\bibitem[{{Bartelmann} {et~al}\mbox{.}(2001){Bartelmann}, {King}, \&
  {Schneider}}]{2001A&A...378..361B}
{Bartelmann} M., {King} L.~J., {Schneider} P., 2001, A\&A, 378, 361

\bibitem[{Bartelmann \& Schneider(2001)}]{Bible}
Bartelmann M., Schneider P., 2001, Physics Reports, 340, 291

\bibitem[{{Ben{\'{\i}}tez}(2000)}]{BPZ}
{Ben{\'{\i}}tez} N., 2000, ApJ, 536, 571

\bibitem[{{Benjamin} {et~al}\mbox{.}(2012){Benjamin}, {Kilbinger}, , {Erben},
  {Kuijken}, {van Waerbeke}, {Heymans}, {Coupon}, {Benjamin}, {Bonnett}, {Fu},
  {Hoekstra}, {Kitching}, \& {Mellier}}]{Benjamin2012}
{Benjamin} J. {et~al.}, 2012, In preparation

\bibitem[{{Benjamin} {et~al}\mbox{.}(2010){Benjamin}, {van Waerbeke},
  {M{\'e}nard}, \& {Kilbinger}}]{JB10}
{Benjamin} J., {van Waerbeke} L., {M{\'e}nard} B., {Kilbinger} M., 2010, MNRAS,
  408, 1168

\bibitem[{Bernstein \& Jarvis(2002)}]{BJ02}
Bernstein G.~M., Jarvis M., 2002, AJ, 123, 583

\bibitem[{{Bertin}(2001)}]{EyE}
{Bertin} E., 2001, in Mining the Sky, {Banday} A.~J., {Zaroubi} S.,
  {Bartelmann} M., eds., p. 353

\bibitem[{{Bertin}(2006)}]{scamp}
{Bertin} E., 2006, in Astronomical Society of the Pacific Conference Series,
  Vol. 351, Astronomical Data Analysis Software and Systems XV, {C.~Gabriel,
  C.~Arviset, D.~Ponz, \& S.~Enrique}, ed., p. 112

\bibitem[{{Bertin} \& {Arnouts}(1996)}]{SExtractor}
{Bertin} E., {Arnouts} S., 1996, A\&AS, 117, 393

\bibitem[{{Bertin} {et~al}\mbox{.}(2002){Bertin}, {Mellier}, {Radovich},
  {Missonnier}, {Didelon}, \& {Morin}}]{swarp}
{Bertin} E., {Mellier} Y., {Radovich} M., {Missonnier} G., {Didelon} P.,
  {Morin} B., 2002, in Astronomical Society of the Pacific Conference Series,
  Vol. 281, Astronomical Data Analysis Software and Systems XI,
  {D.~A.~Bohlender, D.~Durand, \& T.~H.~Handley}, ed., p. 228

\bibitem[{{Bridle et al.}(2010)}]{GREAT08}
{Bridle et al.} S., 2010, MNRAS, 405, 2044

\bibitem[{{Coupon} {et~al}\mbox{.}(2009){Coupon}, {Ilbert}, {Kilbinger},
  {McCracken}, {Mellier}, {Arnouts}, {Bertin}, {Hudelot}, {Schultheis}, {Le
  F{\`e}vre}, {Le Brun}, {Guzzo}, {Bardelli}, {Zucca}, {Bolzonella}, {Garilli},
  {Zamorani}, {Zanichelli}, {Tresse}, \& {Aussel}}]{Coupon09}
{Coupon} J. {et~al.}, 2009, A\&A, 500, 981

\bibitem[{{Cypriano} {et~al}\mbox{.}(2010){Cypriano}, {Amara}, {Voigt},
  {Bridle}, {Abdalla}, {R{\'e}fr{\'e}gier}, {Seiffert}, \&
  {Rhodes}}]{Cypriano10}
{Cypriano} E.~S., {Amara} A., {Voigt} L.~M., {Bridle} S.~L., {Abdalla} F.~B.,
  {R{\'e}fr{\'e}gier} A., {Seiffert} M., {Rhodes} J., 2010, MNRAS, 405, 494

\bibitem[{{Dunkley} {et~al}\mbox{.}(2009){Dunkley}, {Komatsu}, {Nolta},
  {Spergel}, {Larson}, {Hinshaw}, {Page}, {Bennett}, {Gold}, {Jarosik},
  {Weiland}, {Halpern}, {Hill}, {Kogut}, {Limon}, {Meyer}, {Tucker}, {Wollack},
  \& {Wright}}]{WMAP5}
{Dunkley} J. {et~al.}, 2009, ApJS, 180, 306

\bibitem[{{Erben} {et~al}\mbox{.}(2012){Erben}, , {Erben}, {Kuijken}, {van
  Waerbeke}, {Heymans}, {Coupon}, {Benjamin}, {Bonnett}, {Fu}, {Hoekstra},
  {Kitching}, \& {Mellier}}]{Erben12}
{Erben} T. {et~al.}, 2012, In preparation

\bibitem[{{Erben} {et~al}\mbox{.}(2009){Erben}, {Hildebrandt}, {Lerchster},
  {Hudelot}, {Benjamin}, {van Waerbeke}, {Schrabback}, {Brimioulle}, {Cordes},
  {Dietrich}, {Holhjem}, {Schirmer}, \& {Schneider}}]{Erben09}
{Erben} T. {et~al.}, 2009, A\&A, 493, 1197

\bibitem[{{Erben} {et~al}\mbox{.}(2005){Erben}, {Schirmer}, {Dietrich},
  {Cordes}, {Haberzettl}, {Hetterscheidt}, {Hildebrandt}, {Schmithuesen},
  {Schneider}, {Simon}, {Deul}, {Hook}, {Kaiser}, {Radovich}, {Benoist},
  {Nonino}, {Olsen}, {Prandoni}, {Wichmann}, {Zaggia}, {Bomans}, {Dettmar}, \&
  {Miralles}}]{THELI}
{Erben} T. {et~al.}, 2005, Astronomische Nachrichten, 326, 432

\bibitem[{{Erben} {et~al}\mbox{.}(2001){Erben}, {Van Waerbeke}, {Bertin},
  {Mellier}, \& {Schneider}}]{erben}
{Erben} T., {Van Waerbeke} L., {Bertin} E., {Mellier} Y., {Schneider} P., 2001,
  A\&A, 366, 717

\bibitem[{{Ferrarese} {et~al}\mbox{.}(2012){Ferrarese}, {C{\^o}t{\'e}},
  {Cuillandre}, {Gwyn}, {Peng}, {MacArthur}, {Duc}, {Boselli}, {Mei}, {Erben},
  {McConnachie}, {Durrell}, {Mihos}, {Jord{\'a}n}, {Lan{\c c}on}, {Puzia},
  {Emsellem}, {Balogh}, {Blakeslee}, {van Waerbeke}, {Gavazzi}, {Vollmer},
  {Kavelaars}, {Woods}, {Ball}, {Boissier}, {Courteau}, {Ferriere}, {Gavazzi},
  {Hildebrandt}, {Hudelot}, {Huertas-Company}, {Liu}, {McLaughlin}, {Mellier},
  {Milkeraitis}, {Schade}, {Balkowski}, {Bournaud}, {Carlberg}, {Chapman},
  {Hoekstra}, {Peng}, {Sawicki}, {Simard}, {Taylor}, {Tully}, {van Driel},
  {Wilson}, {Burdullis}, {Mahoney}, \& {Manset}}]{NGVS}
{Ferrarese} L. {et~al.}, 2012, ApJS, 200, 4

\bibitem[{{Fu} {et~al}\mbox{.}(2008){Fu}, {Semboloni}, {Hoekstra}, {Kilbinger},
  {van Waerbeke}, {Tereno}, {Mellier}, {Heymans}, {Coupon}, {Benabed},
  {Benjamin}, {Bertin}, {Dor{\'e}}, {Hudson}, {Ilbert}, {Maoli}, {Marmo},
  {McCracken}, \& {M{\'e}nard}}]{Fu08}
{Fu} L. {et~al.}, 2008, A\&A, 479, 9

\bibitem[{{Guy} {et~al}\mbox{.}(2010){Guy}, {Sullivan}, {Conley}, {Regnault},
  {Astier}, {Balland}, {Basa}, {Carlberg}, {Fouchez}, {Hardin}, {Hook},
  {Howell}, {Pain}, {Palanque-Delabrouille}, {Perrett}, {Pritchet}, {Rich},
  {Ruhlmann-Kleider}, {Balam}, {Baumont}, {Ellis}, {Fabbro}, {Fakhouri},
  {Fourmanoit}, {Gonz{\'a}lez-Gait{\'a}n}, {Graham}, {Hsiao}, {Kronborg},
  {Lidman}, {Mourao}, {Perlmutter}, {Ripoche}, {Suzuki}, \& {Walker}}]{SNLS}
{Guy} J. {et~al.}, 2010, A\&A, 523, A7

\bibitem[{{Harnois-D\'{e}raps} {et~al}\mbox{.}(2012){Harnois-D\'{e}raps},
  {Vafaei}, \& {Van Waerbeke}}]{Clone2012}
{Harnois-D\'{e}raps} J., {Vafaei} S., {Van Waerbeke} L., 2012, MNRAS submitted,
  preprint (arXiv:1202.2332)

\bibitem[{Heymans {et~al}\mbox{.}(2005)Heymans, Brown, {Barden}, {Caldwell},
  {Jahnke}, {Rix}, Taylor, {Beckwith}, {Bell}, {Borch}, {H\"au\ss ler},
  {Jogee}, {McIntosh}, {Meisenheimer}, {Peng}, {Sanchez}, {Somerville},
  {Wisotzki}, \& {Wolf}}]{HymzGEMS}
Heymans C. {et~al.}, 2005, MNRAS, 361, 160

\bibitem[{{Heymans} {et~al}\mbox{.}(2008){Heymans}, {Gray}, {Peng}, {van
  Waerbeke}, {Bell}, {Wolf}, {Bacon}, {Balogh}, {Barazza}, {Barden},
  {B{\"o}hm}, {Caldwell}, {H{\"a}u{\ss}ler}, {Jahnke}, {Jogee}, {van Kampen},
  {Lane}, {McIntosh}, {Meisenheimer}, {Mellier}, {S{\'a}nchez}, {Taylor},
  {Wisotzki}, \& {Zheng}}]{Hey08}
{Heymans} C. {et~al.}, 2008, MNRAS, 385, 1431

\bibitem[{{Heymans} {et~al}\mbox{.}(2012){Heymans}, {Rowe}, {Hoekstra},
  {Miller}, {Erben}, {Kitching}, \& {Van Waerbeke}}]{HeyRowe}
{Heymans} C., {Rowe} B., {Hoekstra} H., {Miller} L., {Erben} T., {Kitching} T.,
  {Van Waerbeke} L., 2012, MNRAS, 421, 381

\bibitem[{{Heymans} {et~al}\mbox{.}(2006{\natexlab{a}}){Heymans}, {Van
  Waerbeke}, {Bacon}, {Berge}, {Bernstein}, {Bertin}, {Bridle}, {Brown},
  {Clowe}, {Dahle}, {Erben}, {Gray}, {Hetterscheidt}, {Hoekstra}, {Hudelot},
  {Jarvis}, {Kuijken}, {Margoniner}, {Massey}, {Mellier}, {Nakajima},
  {Refregier}, {Rhodes}, {Schrabback}, \& {Wittman}}]{STEP1}
{Heymans} C. {et~al.}, 2006{\natexlab{a}}, MNRAS, 368, 1323

\bibitem[{{Heymans} {et~al}\mbox{.}(2006{\natexlab{b}}){Heymans}, {White},
  {Heavens}, {Vale}, \& {van Waerbeke}}]{HeymansIA06}
{Heymans} C., {White} M., {Heavens} A., {Vale} C., {van Waerbeke} L.,
  2006{\natexlab{b}}, MNRAS, 371, 750

\bibitem[{{Hildebrandt} {et~al}\mbox{.}(2010){Hildebrandt}, {Arnouts}, {Capak},
  {Moustakas}, {Wolf}, {Abdalla}, {Assef}, {Banerji}, {Ben{\'{\i}}tez},
  {Brammer}, {Budav{\'a}ri}, {Carliles}, {Coe}, {Dahlen}, {Feldmann}, {Gerdes},
  {Gillis}, {Ilbert}, {Kotulla}, {Lahav}, {Li}, {Miralles}, {Purger},
  {Schmidt}, \& {Singal}}]{PHAT}
{Hildebrandt} H. {et~al.}, 2010, A\&A, 523, A31

\bibitem[{{Hildebrandt} {et~al}\mbox{.}(2012){Hildebrandt}, {Erben}, {Kuijken},
  {van Waerbeke}, {Heymans}, {Coupon}, {Benjamin}, {Bonnett}, {Fu}, {Hoekstra},
  {Kitching}, {Mellier}, {Miller}, {Velander}, {Hudson}, {Rowe}, {Schrabback},
  {Semboloni}, \& {Ben{\'{\i}}tez}}]{HH12}
{Hildebrandt} H. {et~al.}, 2012, MNRAS, 421, 2355

\bibitem[{{Hirata} \& {Seljak}(2003)}]{Hirata}
{Hirata} C., {Seljak} U., 2003, MNRAS, 343, 459

\bibitem[{{Hoekstra}(2004)}]{HH04}
{Hoekstra} H., 2004, MNRAS, 347, 1337

\bibitem[{{Hoekstra} {et~al}\mbox{.}(2006){Hoekstra}, {Mellier}, {van
  Waerbeke}, {Semboloni}, {Fu}, {Hudson}, {Parker}, {Tereno}, \&
  {Benabed}}]{HH06}
{Hoekstra} H. {et~al.}, 2006, ApJ, 647, 116

\bibitem[{{Huff} {et~al}\mbox{.}(2011){Huff}, {Eifler}, {Hirata}, {Mandelbaum},
  {Schlegel}, \& {Seljak}}]{Huff}
{Huff} E.~M., {Eifler} T., {Hirata} C.~M., {Mandelbaum} R., {Schlegel} D.,
  {Seljak} U., 2011, MNRAS submitted, preprint (arXiv:1112.3143

\bibitem[{{Ilbert} {et~al}\mbox{.}(2009){Ilbert}, {Capak}, {Salvato}, {Aussel},
  {McCracken}, {Sanders}, {Scoville}, {Kartaltepe}, {Arnouts}, {Le Floc'h},
  {Mobasher}, {Taniguchi}, {Lamareille}, {Leauthaud}, {Sasaki}, {Thompson},
  {Zamojski}, {Zamorani}, {Bardelli}, {Bolzonella}, {Bongiorno}, {Brusa},
  {Caputi}, {Carollo}, {Contini}, {Cook}, {Coppa}, {Cucciati}, {de la Torre},
  {de Ravel}, {Franzetti}, {Garilli}, {Hasinger}, {Iovino}, {Kampczyk},
  {Kneib}, {Knobel}, {Le Borgne}, {Le Brun}, {F{\`e}vre}, {Lilly}, {Looper},
  {Maier}, {Mainieri}, {Mellier}, {Mignoli}, {Murayama}, {Pell{\`o}}, {Peng},
  {P{\'e}rez-Montero}, {Renzini}, {Ricciardelli}, {Schiminovich}, {Scodeggio},
  {Shioya}, {Silverman}, {Surace}, {Tanaka}, {Tasca}, {Tresse}, {Vergani}, \&
  {Zucca}}]{COSMOS30}
{Ilbert} O. {et~al.}, 2009, ApJ, 690, 1236

\bibitem[{{Joachimi} {et~al}\mbox{.}(2011){Joachimi}, {Mandelbaum}, {Abdalla},
  \& {Bridle}}]{BJ11}
{Joachimi} B., {Mandelbaum} R., {Abdalla} F.~B., {Bridle} S.~L., 2011, A\&A,
  527, A26

\bibitem[{{Kacprzak} {et~al}\mbox{.}(2012){Kacprzak}, {Zuntz}, {Rowe},
  {Bridle}, {Refregier}, {Amara}, {Voigt}, \& {Hirsch}}]{Kacprzak}
{Kacprzak} T., {Zuntz} J., {Rowe} B., {Bridle} S., {Refregier} A., {Amara} A.,
  {Voigt} L., {Hirsch} M., 2012, MNRAS submitted, preprint (arXiv:1203.5049)

\bibitem[{Kaiser(2000)}]{K00}
Kaiser N., 2000, ApJ, 537, 555

\bibitem[{Kaiser {et~al}\mbox{.}(1995)Kaiser, Squires, \& Broadhurst}]{KSB}
Kaiser N., Squires G., Broadhurst T., 1995, ApJ, 449, 460

\bibitem[{{Kilbinger} {et~al}\mbox{.}(2012){Kilbinger}, , {Erben}, {Kuijken},
  {van Waerbeke}, {Heymans}, {Coupon}, {Benjamin}, {Bonnett}, {Fu}, {Hoekstra},
  {Kitching}, \& {Mellier}}]{Kilbinger2012}
{Kilbinger} M. {et~al.}, 2012, MNRAS submitted

\bibitem[{{Kilbinger} {et~al}\mbox{.}(2009){Kilbinger}, {Benabed}, {Guy},
  {Astier}, {Tereno}, {Fu}, {Wraith}, {Coupon}, {Mellier}, {Balland},
  {Bouchet}, {Hamana}, {Hardin}, {McCracken}, {Pain}, {Regnault}, {Schultheis},
  \& {Yahagi}}]{MK09}
{Kilbinger} M. {et~al.}, 2009, A\&A, 497, 677

\bibitem[{{Kilbinger} {et~al}\mbox{.}(2006){Kilbinger}, {Schneider}, \&
  {Eifler}}]{KSE06}
{Kilbinger} M., {Schneider} P., {Eifler} T., 2006, A\&A, 457, 15

\bibitem[{{Kitching} {et~al}\mbox{.}(2012{\natexlab{a}}){Kitching}, {Balan},
  {Bridle}, {Cantale}, {Courbin}, {Gentile}, {Gill}, {Harmeling}, {Heymans},
  {Hirsch}, {Kacprzak}, {Kirkby}, {Margala}, {Massey}, {Melchior}, {Nurbaeva},
  {Patton}, {Rhodes}, {Rowe}, {Taylor}, {Tewes}, {Viola}, {Witherick}, {Voigt},
  {Young}, \& {Zuntz}}]{GREAT10}
{Kitching} T.~D. {et~al.}, 2012{\natexlab{a}}, MNRAS, 423, 3163

\bibitem[{{Kitching} {et~al}\mbox{.}(2008){Kitching}, {Miller}, {Heymans}, {van
  Waerbeke}, \& {Heavens}}]{Lensfit2}
{Kitching} T.~D., {Miller} L., {Heymans} C.~E., {van Waerbeke} L., {Heavens}
  A.~F., 2008, MNRAS, 390, 149

\bibitem[{{Kitching} {et~al}\mbox{.}(2012{\natexlab{b}}){Kitching}, {Rhodes},
  {Heymans}, {Massey}, {Liu}, {Cobzarenco}, {Cragin}, {Hassaine}, {Kirkby},
  {Lok}, {Margala}, {Moser}, {O'Leary}, {Pires}, \& {Yurgenson}}]{MDM}
{Kitching} T.~D. {et~al.}, 2012{\natexlab{b}}, New Astronomy Reviews submitted,
  preprint (arXiv:1204.4096)

\bibitem[{{Kuijken}(2006)}]{KKshapes}
{Kuijken} K., 2006, A\&A, 456, 827

\bibitem[{{Le F{\`e}vre} {et~al}\mbox{.}(2005){Le F{\`e}vre}, {Vettolani},
  {Garilli}, {Tresse}, {Bottini}, {Le Brun}, {Maccagni}, {Picat}, {Scaramella},
  {Scodeggio}, {Zanichelli}, {Adami}, {Arnaboldi}, {Arnouts}, {Bardelli},
  {Bolzonella}, {Cappi}, {Charlot}, {Ciliegi}, {Contini}, {Foucaud},
  {Franzetti}, {Gavignaud}, {Guzzo}, {Ilbert}, {Iovino}, {McCracken}, {Marano},
  {Marinoni}, {Mathez}, {Mazure}, {Meneux}, {Merighi}, {Paltani}, {Pell{\`o}},
  {Pollo}, {Pozzetti}, {Radovich}, {Zamorani}, {Zucca}, {Bondi}, {Bongiorno},
  {Busarello}, {Lamareille}, {Mellier}, {Merluzzi}, {Ripepi}, \&
  {Rizzo}}]{VVDS05}
{Le F{\`e}vre} O. {et~al.}, 2005, A\&A, 439, 845

\bibitem[{{Magnier} \& {Cuillandre}(2004)}]{Elixir}
{Magnier} E.~A., {Cuillandre} J.-C., 2004, Publ. Astron. Soc. Pacific, 116, 449

\bibitem[{{Mandelbaum} {et~al}\mbox{.}(2012){Mandelbaum}, {Hirata},
  {Leauthaud}, {Massey}, \& {Rhodes}}]{SHERA}
{Mandelbaum} R., {Hirata} C.~M., {Leauthaud} A., {Massey} R.~J., {Rhodes} J.,
  2012, MNRAS, 420, 1518

\bibitem[{{Massey} {et~al}\mbox{.}(2007{\natexlab{a}}){Massey}, {Heymans},
  {Berg{\'e}}, {Bernstein}, {Bridle}, {Clowe}, {Dahle}, {Ellis}, {Erben},
  {Hetterscheidt}, {High}, {Hirata}, {Hoekstra}, {Hudelot}, {Jarvis},
  {Johnston}, {Kuijken}, {Margoniner}, {Mandelbaum}, {Mellier}, {Nakajima},
  {Paulin-Henriksson}, {Peeples}, {Roat}, {Refregier}, {Rhodes}, {Schrabback},
  {Schirmer}, {Seljak}, {Semboloni}, \& {van Waerbeke}}]{STEP2}
{Massey} R. {et~al.}, 2007{\natexlab{a}}, MNRAS, 376, 13

\bibitem[{{Massey} {et~al}\mbox{.}(2007{\natexlab{b}}){Massey}, {Rhodes},
  {Ellis}, {Scoville}, {Leauthaud}, {Finoguenov}, {Capak}, {Bacon}, {Aussel},
  {Kneib}, {Koekemoer}, {McCracken}, {Mobasher}, {Pires}, {Refregier},
  {Sasaki}, {Starck}, {Taniguchi}, {Taylor}, \& {Taylor}}]{MasseyNat}
{Massey} R. {et~al.}, 2007{\natexlab{b}}, Nature, 445, 286

\bibitem[{{Massey} {et~al}\mbox{.}(2007{\natexlab{c}}){Massey}, {Rowe},
  {Refregier}, {Bacon}, \& {Berg{\'e}}}]{shapelets07}
{Massey} R., {Rowe} B., {Refregier} A., {Bacon} D.~J., {Berg{\'e}} J.,
  2007{\natexlab{c}}, MNRAS, 380, 229

\bibitem[{{Melchior} \& {Viola}(2012)}]{MelchiorViola}
{Melchior} P., {Viola} M., 2012, MNRAS, 424, 2757

\bibitem[{{Miller} {et~al}\mbox{.}(2012){Miller}, {Erben}, {Kuijken}, {van
  Waerbeke}, {Heymans}, {Coupon}, {Benjamin}, {Bonnett}, {Fu}, {Hoekstra},
  {Kitching}, \& {Mellier}}]{Miller2012}
{Miller} L. {et~al.}, 2012, MNRAS submitted

\bibitem[{{Miller} {et~al}\mbox{.}(2007){Miller}, {Kitching}, {Heymans},
  {Heavens}, \& {van Waerbeke}}]{Lensfit1}
{Miller} L., {Kitching} T.~D., {Heymans} C., {Heavens} A.~F., {van Waerbeke}
  L., 2007, MNRAS, 382, 315

\bibitem[{{Moffat}(1969)}]{Moffat}
{Moffat} A.~F.~J., 1969, A\&A, 3, 455

\bibitem[{{Newman} {et~al}\mbox{.}(2012){Newman}, {Cooper}, {Davis}, {Faber},
  {Coil}, {Guhathakurta}, {Koo}, {Phillips}, {Conroy}, {Dutton}, {Finkbeiner},
  {Gerke}, {Rosario}, {Weiner}, {Willmer}, {Yan}, {Harker}, {Kassin},
  {Konidaris}, {Lai}, {Madgwick}, {Noeske}, {Wirth}, {Connolly}, {Kaiser},
  {Kirby}, {Lemaux}, {Lin}, {Lotz}, {Luppino}, {Marinoni}, {Matthews},
  {Metevier}, \& {Schiavon}}]{DEEP2}
{Newman} J.~A. {et~al.}, 2012, ApJS submitted, preprint (arXiv:1203.3192)

\bibitem[{Press {et~al}\mbox{.}(1986)Press, Teukolsky, Vetterling, \&
  Flannery}]{Numrec}
Press W., Teukolsky S., Vetterling W., Flannery B., 1986, Numerical recipes.
  Cambridge University Press

\bibitem[{{Refregier} {et~al}\mbox{.}(2012){Refregier}, {Kacprzak}, {Amara},
  {Bridle}, \& {Rowe}}]{RefKac}
{Refregier} A., {Kacprzak} T., {Amara} A., {Bridle} S., {Rowe} B., 2012, MNRAS,
  425, 1951

\bibitem[{{Reyes} {et~al}\mbox{.}(2010){Reyes}, {Mandelbaum}, {Seljak},
  {Baldauf}, {Gunn}, {Lombriser}, \& {Smith}}]{Reyes}
{Reyes} R., {Mandelbaum} R., {Seljak} U., {Baldauf} T., {Gunn} J.~E.,
  {Lombriser} L., {Smith} R.~E., 2010, Nature, 464, 256

\bibitem[{{Rhodes} {et~al}\mbox{.}(2007){Rhodes}, {Massey}, {Albert},
  {Collins}, {Ellis}, {Heymans}, {Gardner}, {Kneib}, {Koekemoer}, {Leauthaud},
  {Mellier}, {Refregier}, {Taylor}, \& {Van Waerbeke}}]{RhodesPSF}
{Rhodes} J.~D. {et~al.}, 2007, ApJS, 172, 203

\bibitem[{{Rowe}(2010)}]{Rowe10}
{Rowe} B., 2010, MNRAS, 404, 350

\bibitem[{{Schneider} {et~al}\mbox{.}(2010){Schneider}, {Eifler}, \&
  {Krause}}]{COSEBIS}
{Schneider} P., {Eifler} T., {Krause} E., 2010, A\&A, 520, A116

\bibitem[{{Schneider} \& {Kilbinger}(2007)}]{SK07}
{Schneider} P., {Kilbinger} M., 2007, A\&A, 462, 841

\bibitem[{{Schneider} {et~al}\mbox{.}(2002){Schneider}, {Van Waerbeke},
  {Kilbinger}, \& {Mellier}}]{SchvWKM02}
{Schneider} P., {Van Waerbeke} L., {Kilbinger} M., {Mellier} Y., 2002, A\&A,
  369, 1

\bibitem[{{Schrabback} {et~al}\mbox{.}(2010){Schrabback}, {Hartlap},
  {Joachimi}, {Kilbinger}, {Simon}, {Benabed}, {Brada{\v c}}, {Eifler},
  {Erben}, {Fassnacht}, {High}, {Hilbert}, {Hildebrandt}, {Hoekstra},
  {Kuijken}, {Marshall}, {Mellier}, {Morganson}, {Schneider}, {Semboloni}, {van
  Waerbeke}, \& {Velander}}]{TS10}
{Schrabback} T. {et~al.}, 2010, A\&A, 516, A63

\bibitem[{{Shao}(1986)}]{1986ANNSTAT}
{Shao} J., 1986, Ann. Stat., 14, 1322

\bibitem[{{Simpson} {et~al}\mbox{.}(2012){Simpson}, , {Erben}, {Kuijken}, {van
  Waerbeke}, {Heymans}, {Coupon}, {Benjamin}, {Bonnett}, {Fu}, {Hoekstra},
  {Kitching}, \& {Mellier}}]{Simpson}
{Simpson} F. {et~al.}, 2012, MNRAS submitted

\bibitem[{{van Uitert} {et~al}\mbox{.}(2011){van Uitert}, {Hoekstra},
  {Velander}, {Gilbank}, {Gladders}, \& {Yee}}]{RCS2}
{van Uitert} E., {Hoekstra} H., {Velander} M., {Gilbank} D.~G., {Gladders}
  M.~D., {Yee} H.~K.~C., 2011, A\&A, 534, A14

\bibitem[{{Van Waerbeke} {et~al}\mbox{.}(2000){Van Waerbeke}, {Mellier},
  {Erben}, {Cuillandre}, {Bernardeau}, {Maoli}, {Bertin}, {Mc Cracken}, {Le
  F\`evre}, {Fort}, {Dantel-Fort}, {Jain}, \& {Schneider}}]{vWb00}
{Van Waerbeke} L. {et~al.}, 2000, A\&A, 358, 30

\bibitem[{{Van Waerbeke} {et~al}\mbox{.}(2005){Van Waerbeke}, {Mellier}, \&
  {Hoekstra}}]{vWb05}
{Van Waerbeke} L., {Mellier} Y., {Hoekstra} H., 2005, A\&A, 429, 75

\bibitem[{{Velander} {et~al}\mbox{.}(2012){Velander}, , {Erben}, {Kuijken},
  {van Waerbeke}, {Heymans}, {Coupon}, {Benjamin}, {Bonnett}, {Fu}, {Hoekstra},
  {Kitching}, \& {Mellier}}]{Velander2012}
{Velander} M. {et~al.}, 2012, MNRAS submitted

\bibitem[{{Velander} {et~al}\mbox{.}(2011){Velander}, {Kuijken}, \&
  {Schrabback}}]{Velander}
{Velander} M., {Kuijken} K., {Schrabback} T., 2011, MNRAS, 412, 2665

\bibitem[{{Voigt} {et~al}\mbox{.}(2012){Voigt}, {Bridle}, {Amara}, {Cropper},
  {Kitching}, {Massey}, {Rhodes}, \& {Schrabback}}]{Voigt}
{Voigt} L.~M., {Bridle} S.~L., {Amara} A., {Cropper} M., {Kitching} T.~D.,
  {Massey} R., {Rhodes} J., {Schrabback} T., 2012, MNRAS, 421, 1385

\bibitem[{{Weinberg} {et~al}\mbox{.}(2012){Weinberg}, Mortonson, Eisenstein,
  Hirata, Riess, \& Rozo}]{Weinberg}
{Weinberg} D., Mortonson M., Eisenstein D., Hirata C., Riess A., Rozo E., 2012,
  Physics Reports

\bibitem[{Wittman {et~al}\mbox{.}(2000)Wittman, Tyson, Kirkman, Bernstein, \&
  Dell'Antonio}]{Witt}
Wittman D., Tyson J., Kirkman D., Bernstein G., Dell'Antonio I., 2000, Nature,
  405, 143

\end{thebibliography}
\label{lastpage}

\end{document}